\documentclass[journal]{IEEEtran}
\IEEEoverridecommandlockouts
\usepackage{kotex}
\usepackage{cite}
\usepackage{amsmath,amssymb,amsfonts}
\usepackage{algorithm}
\usepackage{algpseudocode}
\usepackage{graphicx}
\usepackage{textcomp}
\usepackage{xcolor}
\usepackage{dsfont}
\usepackage{bbm}
\usepackage[caption=false,font=footnotesize]{subfig}

\def\BibTeX{{\rm B\kern-.05em{\sc i\kern-.025em b}\kern-.08em
    T\kern-.1667em\lower.7ex\hbox{E}\kern-.125emX}}

\begin{document}
\title{Prediction-Aided V2X Safety Message Recovery via Uncertainty-Aware LDPC Decoding}

\author{Sojeong Park, Hyeonsu Lyu, Minwoo Kim, and Hyun Jong Yang,~\IEEEmembership{Senior Member,~IEEE}

\thanks{
S. Park and M. Kim are with the Department of Electrical Engineering, Pohang University of Science and Technology (POSTECH), Pohang, Korea (e-mail: \{sojeong, mwkim0210\}@postech.ac.kr).
H. Lyu is with the Institute of New Media and Communications, Seoul National University, Seoul, Korea (e-mail: hs.lyu@snu.ac.kr).
H. J. Yang is with the Department of Electrical and Computer Engineering and the Institute of New Media and Communications, Seoul National University, Seoul, Korea (e-mail: hjyang@snu.ac.kr).
H. J. Yang is the corresponding author.
}
}
\maketitle

\begin{abstract}
Periodic basic safety messages (BSMs) exhibit temporal correlation as vehicle motion evolves continuously over time. This temporal structure provides predictive information about the current BSM and motivates prediction-aided recovery after a decoding failure. For low-density parity-check (LDPC)-coded transmission, a straightforward approach maps a predicted vehicle state to a candidate bit sequence and uses it as decoder prior information. However, a point prediction does not capture the uncertainty of the prediction. The resulting hard point prior assigns fixed confidence to each predicted bit and may introduce strong erroneous evidence when the prediction is incorrect. This paper proposes an uncertainty-aware prediction-aided LDPC recovery framework for vehicle-to-everything (V2X) safety messages. Instead of using a deterministic point prediction, the proposed method represents the current vehicle state by a predictive distribution. The distribution is propagated through BSM field quantization and serialization to obtain bit-level probabilities, which are converted into prior log-likelihood ratios (LLRs) whose magnitudes reflect prediction reliability. After an initial cyclic redundancy check (CRC) failure, these probabilistic priors are combined with the original channel observations and used in a second LDPC decoding pass. The proposed method preserves the standardized message representation and channel-coding procedure. Simulation results show that the proposed probabilistic priors outperform the hard point prior across different motion predictors. At $E_b/N_0=0.75$~dB, the proposed method recovers 87.27\% of the messages that fail the initial decoding attempt.
\end{abstract}


\section{Introduction}

Autonomous driving has gained substantial attention in recent years, with extensive research advancing vehicle perception, motion prediction, and automated decision-making~\cite{atakishiyev2024explainable, gomezhuelamo2024efficient, shao2023failure, zabolotnii2025pedestrian, prutsch2024efficient, dalcol20205joint}. Safe operation of autonomous vehicles relies on timely and reliable information about surrounding road users and their evolving motion states~\cite{zhang2020sensing}. Although onboard sensors provide direct observations of the driving environment, their coverage can be limited by occlusions and sensing range. Vehicle-to-everything (V2X) communication complements onboard sensing by enabling vehicles to exchange motion and status information with nearby vehicles and roadside infrastructure~\cite{rishiwal2024vehicle}. Recognizing the role of such information exchange, 3GPP specifies service requirements for advanced driving, vehicle platooning, and extended sensor sharing in TS~22.186~\cite{3gpp22186}, including latency and reliability requirements for these applications. Meeting these requirements is important for maintaining timely awareness of surrounding traffic and supporting cooperative driving decisions.

The basic safety message (BSM), defined in SAE J2735~\cite{sae2024j2735}, supports this information exchange by carrying vehicle position, speed, heading, acceleration, and other safety-relevant data. Through periodic BSM broadcasts, each vehicle shares its own motion state with nearby vehicles, allowing receiving vehicles to continuously track the motion states of surrounding traffic. A lost message extends the interval between usable state updates, leaving the receiver to rely on older observations. In the new radio (NR) sidelink shared channel (SL-SCH) coding chain, a cyclic redundancy check (CRC) is used to detect residual errors after low-density parity-check (LDPC) decoding~\cite{ts38212rel17}. A CRC failure indicates that residual errors remain in the decoded transport block (TB), making the block unusable.
Consequently, the receiver must discard the failed TB and wait for the next successfully decoded transmission before obtaining an updated vehicle state.
Therefore, recovering a failed TB is important for maintaining timely vehicle-state updates without waiting for the next message.
The temporally correlated vehicle states in previously received BSMs can be used to recover the motion-related fields of a failed message.
Since vehicle motion evolves over time, previously received BSMs provide additional information to decode the current message \cite{Lu23-TITS, Chang24-TVT}.

\begin{figure} [t]
    \centering
    \includegraphics[width=0.97\linewidth]{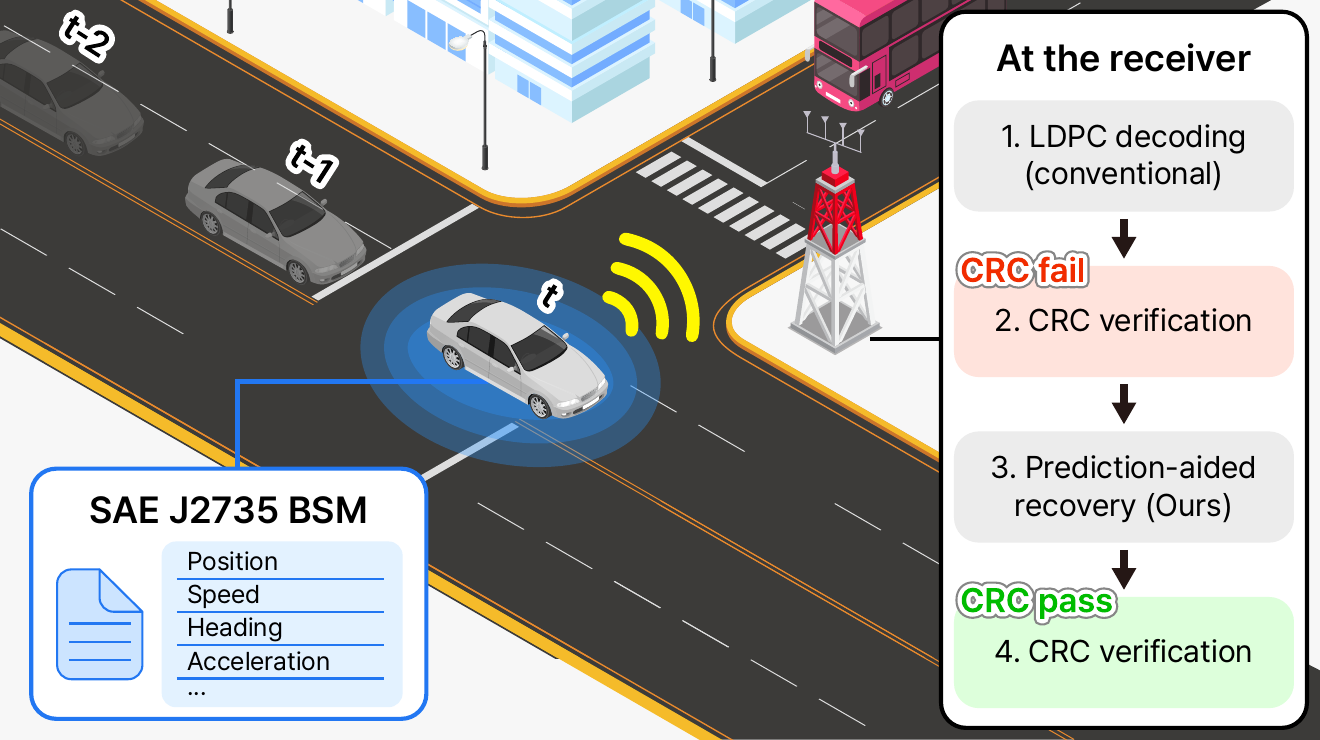}
    \caption{Illustration of the proposed prediction-aided recovery framework for SAE J2735 V2X safety messages.}
    \label{fig:scenario}
\end{figure}

This predictability suggests that the application-level content of BSMs can assist the recovery of their transmitted bits. The idea is related to semantic and task-oriented communication, which incorporates source content, context, or task relevance into communication-system design~\cite{gunduz2023beyond,luo2022semantic,guo2024survey}. 
Related studies have leveraged source characteristics, semantic information, and task relevance to improve communication efficiency and reliability~\cite{bourtsoulatze2019deep,xie2021deep,park2026robust,liu2024adaptable, noh2024drl}. From a channel-decoding perspective, this idea can be viewed more specifically as exploiting application-level side information to construct a priori information for the decoder.
In the present setting, the relevant context is the temporal evolution of vehicle motion, and the objective is the recovery of the original standardized BSM. However, motion prediction does not directly provide bit-level information for LDPC decoding. A motion predictor operates in a continuous state domain, whereas an LDPC decoder operates on the binary representation produced by field quantization and serialization. Even a small state-prediction error can change the resulting bit pattern. Moreover, a point prediction yields only a single BSM candidate, discarding the uncertainty associated with the predicted state. Therefore, a probabilistic mapping is required to translate vehicle-state prediction and its uncertainty into bit-level soft information.

The use of source-side information to assist channel decoding has a substantial history. Source-controlled decoding incorporates source-bit probabilities and residual source correlation as a priori soft information for channel decoding~\cite{hagenauer1995source}. Joint source-channel decoding further extends this idea to LDPC-coded correlated sources, where temporal memory and inter-source correlation provide additional information for iterative decoding~\cite{asvadi2013joint,khas2018design}. More recently, semantic information also enters the soft-decoding domain for text transmission. LLM-based receivers use reconstruction confidence or parity-verified semantic corrections to construct soft priors for subsequent LDPC decoding~\cite{thai2026soft,park2026semanticldpc}. These studies demonstrate the value of side information beyond the instantaneous channel observation. Unlike these approaches, this work derives decoder side information from a predictive distribution of the current vehicle state inferred from previously recovered BSMs. The key distinction is how uncertainty in the predicted physical state is incorporated into the decoding process.

In this paper, we propose an uncertainty-aware prediction-aided LDPC recovery framework for V2X safety messages. Rather than converting a single predicted vehicle state into a fixed-confidence bit prior, the proposed framework exploits the temporal correlation of consecutive BSMs to construct uncertainty-aware bit-level soft information for LDPC decoding. After an initial CRC failure, the receiver predicts the current vehicle state as a distribution from previously recovered BSMs. We propagate this predictive distribution through BSM field quantization and serialization to obtain probabilistic bit-level priors, whose LLR magnitudes reflect the uncertainty of the underlying motion prediction. These priors are combined with the original channel observations and used for a second LDPC decoding pass. In this way, the proposed framework translates uncertainty in continuous-domain vehicle-state prediction into bit-level soft information for channel decoding. Importantly, this process does not require any modification to either the transmitted message format or the standardized channel-coding chain.
For a standards-aligned implementation, we consider SAE J2735 BSM Core Data Part I with ASN.1 unaligned Packed Encoding Rules (UPER) serialization~\cite{sae2024j2735,itut2021x691}.
The channel-coding procedure of SL-SCH follows 3GPP TS~38.212~\cite{ts38212rel17}, while resource-related parameters are derived from a TS~38.533 reference configuration~\cite{ts38533rel17}.
The main contributions of this work are summarized as follows:
\begin{itemize}
    \item
    We propose a prediction-aided LDPC recovery framework for periodic V2X safety messages. 
    When the initial decoding attempt fails CRC verification, temporal information from previously recovered BSMs is exploited to assist the recovery of the current TB, without requiring retransmission or additional channel observations.

    \item
    We develop an uncertainty-aware bit-prior construction that addresses the mismatch between continuous-domain motion prediction and bit-level channel decoding.
    A small error in the predicted vehicle state can change the serialized bit pattern after quantization and serialization. Moreover, a point prediction does not capture the uncertainty of the prediction.
    To address both issues, we propagate the predictive vehicle-state distribution through BSM field quantization and serialization to obtain bit-level probabilities and prior LLRs whose magnitudes reflect prediction reliability.

    \item
    We characterize the effect of prediction uncertainty on decoder priors by comparing the proposed probabilistic prior with a hard point prior.
    The analysis shows that uncertainty-aware LLRs reduce the influence of unreliable predictions, whereas a hard point prior can introduce strong conflicting evidence when its predicted bit is incorrect.
    The probabilistic priors are incorporated into a second LDPC decoding pass while preserving the original channel observations, transmitted BSM representation, and underlying channel-coding procedure.
\end{itemize}

The remainder of this paper is organized as follows. Section~\ref{sec:system_model} introduces the V2X transmission model. Section~\ref{sec:pred} presents the proposed prediction-aided LDPC recovery framework. Section~\ref{sec:results} describes the simulation setup and presents the performance results. Section~\ref{sec:conclusion} concludes the paper.

\section{System Model} \label{sec:system_model}

\begin{figure*}[t]
    \centering
    \includegraphics[width=18cm]{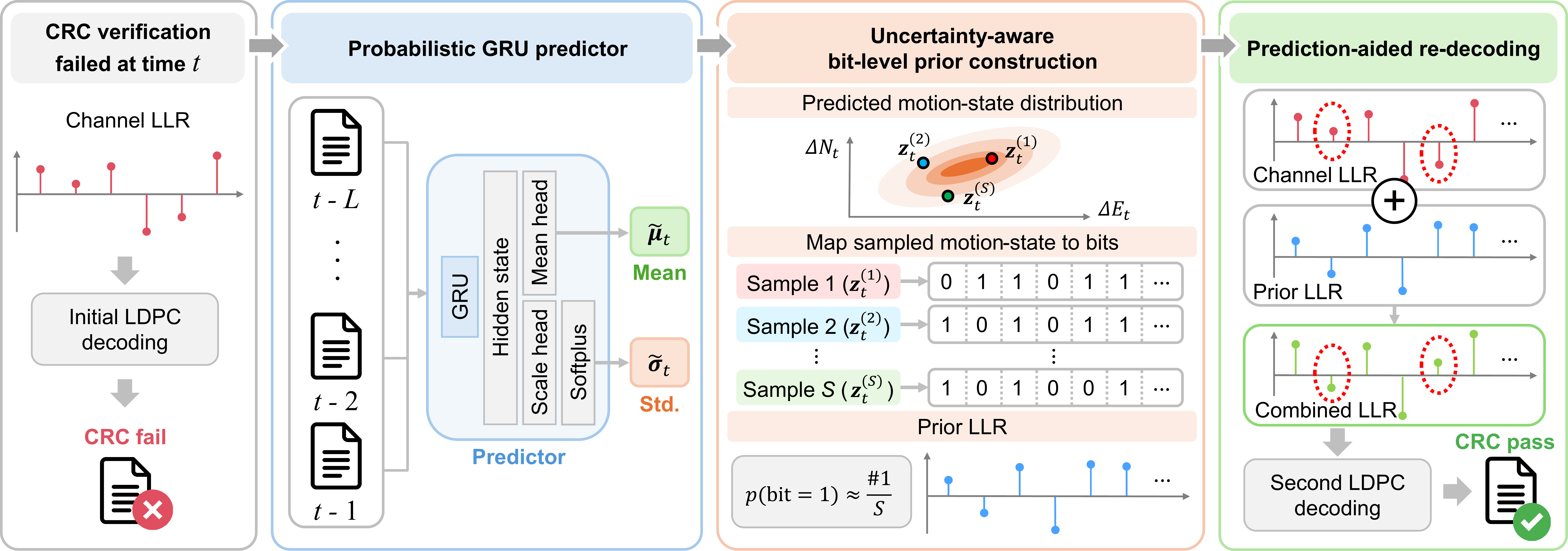}
    \caption{The proposed prediction-aided LDPC re-decoding framework using a probabilistic GRU predictor and uncertainty-aware bit-level priors.}
    \label{fig:system_model}
\end{figure*}

We consider a V2X safety-message transmission system in which a vehicle periodically broadcasts an SAE J2735 BSM describing its motion state. At each transmission instant, the serialized BSM is incorporated into a TB, which is protected by a CRC and LDPC-encoded before transmission over a noisy wireless channel.
At the receiver, conventional LDPC decoding and CRC verification are first performed. Successfully verified TBs are accepted directly. If CRC verification fails, the proposed recovery procedure is applied. The receiver uses a history of previous BSMs to obtain a probabilistic prediction of the vehicle state and exploits this information to assist LDPC re-decoding.

\subsection{V2X Transmission Model}

For the message representation, we consider the SAE J2735 BSM Core Data Part I and serialize it according to the ASN.1 UPER. The serialized BSM generated at transmission instant $t$ is represented by
\begin{equation}
\mathbf{m}_t \in \{0,1\}^{M},
\end{equation}
where $M$ denotes the serialized BSM length.

Among the serialized BSM bits, a subset corresponds to temporally predictable motion-related fields, including position and speed. We denote their bit positions by
\begin{equation}
\mathcal{I}_{\mathrm{pred}} \subseteq \{1,\ldots,M\}.
\end{equation}

To match the TB size of the considered NR sidelink configuration, the serialized BSM is concatenated with a deterministic padding sequence
$\mathbf{q}_t \in \{0,1\}^{L_{\mathrm{pad}}}$.
The resulting TB payload is
\begin{equation}
\mathbf{a}_t = \left[\mathbf{m}_t,\, \mathbf{q}_t \right] \in \{0,1\}^{A},
\; A=M+L_{\mathrm{pad}}.
\label{eq:tb_payload}
\end{equation}

An $L_{\mathrm{CRC}}$-bit CRC is computed over the entire TB as
\begin{equation}
\mathbf{r}_t = f_{\mathrm{CRC}}(\mathbf{a}_t) \in \{0,1\}^{L_{\mathrm{CRC}}},
\label{eq:tb_crc}
\end{equation}
where $f_{\mathrm{CRC}}(\cdot)$ denotes the CRC encoding operation. After CRC attachment, the sequence presented to the subsequent channel coding stage is
\begin{equation}
\mathbf{b}_t = \left[\mathbf{a}_t,\,\mathbf{r}_t\right] =
\left[\mathbf{m}_t,\, \mathbf{q}_t,\, \mathbf{r}_t \right]
\in \{0,1\}^{B}, \; B=A+L_{\mathrm{CRC}}.
\label{eq:crc_appended_block}
\end{equation}

The CRC-appended TB is then processed by the LDPC coding chain. Let
\begin{equation}
\mathbf{v}_t = f_{\mathrm{LDPC}}(\mathbf{b}_t)
\end{equation}
denote the corresponding LDPC mother-code representation. The mother-code
bits are subsequently rate matched according to the selected sidelink
transmission configuration
\begin{equation}
\mathbf{c}_t = f_{\mathrm{RM}}(\mathbf{v}_t) \in \{0,1\}^{E},
\label{eq:rate_matched_bits}
\end{equation}
where $f_{\mathrm{RM}}(\cdot)$ denotes the NR rate-matching operation and $E$ is the number of rate-matched coded bits transmitted over the wireless channel.

\subsection{Initial Decoding and CRC Verification}

At the receiver, the channel observations corresponding to the  rate-matched coded bits are converted into bit-wise channel log-likelihood ratios (LLRs) without using any prediction-derived information. Let 
\begin{equation} 
\mathbf{L}_{\mathrm{ch},t} \in \mathbb{R}^{E} 
\end{equation} 
denote the channel LLR vector associated with $\mathbf{c}_t$, whose  $k$-th element is defined as 
\begin{equation} 
L_{\mathrm{ch},t}(k) = 
\log \frac{p(\mathbf{y}_t \mid c_{t,k}=1)}{p(\mathbf{y}_t \mid c_{t,k}=0)}, 
\; k=1,\ldots,E, 
\end{equation} 
where $\mathbf{y}_t$ denotes the received channel observations. After mapping the channel LLRs to the LDPC mother-code domain through NR de-rate matching, belief propagation (BP) decoding is performed with a maximum of $I_{\max}$ iterations.

We denote the resulting estimate of the CRC-appended TB by
\begin{equation}
\hat{\mathbf{b}}_{t}^{(0)} = \mathcal{D}_{\mathrm{BP}}
\left( f_{\mathrm{DRM}}(\mathbf{L}_{\mathrm{ch},t}); I_{\max} \right)
= \left[ \hat{\mathbf{a}}_{t}^{(0)}, \hat{\mathbf{r}}_{t}^{(0)}\right],
\label{eq:initial_bp}
\end{equation}
where $\hat{\mathbf{a}}_{t}^{(0)}$ and $\hat{\mathbf{r}}_{t}^{(0)}$ denote the estimated TB payload and its CRC bits, respectively. The operator $f_{\mathrm{DRM}}(\cdot)$ maps the received channel LLRs to the LDPC variable-node domain through de-rate matching, including the treatment of punctured and filler positions. The operator $\mathcal{D}_{\mathrm{BP}}(\cdot;\cdot)$ takes LLRs in this domain and performs BP message passing, hard-decision extraction, and recovery of the CRC-appended TB, excluding filler bits. Its second argument specifies the number of BP iterations. The recovered TB is further expressed as
\begin{equation}
\hat{\mathbf{a}}_{t}^{(0)} =
\left[\hat{\mathbf{m}}_{t}^{(0)}, \hat{\mathbf{q}}_{t}^{(0)} \right]
\end{equation}
where $\hat{\mathbf{m}}_{t}^{(0)}$ and $\hat{\mathbf{q}}_{t}^{(0)}$ denote the recovered BSM and deterministic padding sequences, respectively. The superscript $(0)$ indicates quantities obtained from the initial decoding pass.

CRC verification is then performed over the recovered TB. Specifically, the CRC is recomputed from $\hat{\mathbf{a}}_{t}^{(0)}$ and compared with the decoded CRC bits $\hat{\mathbf{r}}_{t}^{(0)}$
\begin{equation}
\eta_t =
\begin{cases}
1, & f_{\mathrm{CRC}} \left( \hat{\mathbf{a}}_{t}^{(0)} \right) = \hat{\mathbf{r}}_{t}^{(0)},
\\
0, & \text{otherwise}.
\end{cases}
\label{eq:initial_crc}
\end{equation}
$\eta_t=1$ indicates that the recovered TB passes the CRC, whereas $\eta_t=0$ indicates an initial decoding failure. TBs that pass the CRC are accepted without further processing. Only TBs that fail the CRC enter the prediction-aided recovery stage. For these failed TBs, the same channel LLRs are reused together with prediction-derived side information to perform a fresh BP decoding pass. No retransmission or additional channel observation is required.

\subsection{Predictive Side Information}
\label{subsec:predictive_side_info}

Vehicle motion changes smoothly over consecutive BSM transmissions. Previously recovered BSMs provide useful information for predicting the position and speed represented in the current BSM. The receiver uses this prediction to construct side information when the current TB fails initial decoding.
Let
\begin{equation}
\mathcal{H}_t = \left\{ \mathbf{x}_{t-L},\ldots,\mathbf{x}_{t-1} \right\}
\end{equation}
denote a history of $L$ previously recovered BSMs preceding transmission instant $t$, where $\mathbf{x}_{\tau}$ represents the vehicle-state information extracted from the BSM at time $\tau$. The BSMs included in $\mathcal{H}_t$ are assumed to have been reliably recovered at the receiver.

The motion state to be predicted at transmission instant $t$ is represented by
\begin{equation}
\mathbf{z}_t = \left[\Delta E_t,\,\Delta N_t,\,v_t\right]^{\mathsf T},
\end{equation}
where $\Delta E_t$ and $\Delta N_t$ denote the east and north displacements relative to the vehicle position at time $t-1$, and $v_t$ denotes the vehicle speed.
Rather than producing only a point estimate, the predictor models the conditional distribution of the target motion state given the available history. We denote the resulting predictive distribution by
\begin{equation}
\mathcal{P}_t = p\left(\mathbf{z}_t \mid \mathcal{H}_t\right)
= f_{\mathrm{pred}}(\mathcal{H}_t),
\label{eq:predictive_distribution}
\end{equation}
where $f_{\mathrm{pred}}(\cdot)$ represents the probabilistic prediction
model.

When the TB at transmission instant $t$ fails the initial CRC verification, the receiver uses $\mathcal{P}_t$ to estimate the predictable fields of the corresponding BSM in probabilistic form. These field-level predictions are translated into soft bit priors for the serialized BSM positions in $\mathcal{I}_{\mathrm{pred}}$ and mapped to the associated systematic LDPC variables.
The prediction-derived priors are then combined with the original channel information and used in a second LDPC decoding pass. 
Further details are provided in Section~\ref{sec:pred}.

\section{Prediction-Aided LDPC Recovery} \label{sec:pred}

This section presents the proposed prediction-aided recovery scheme for TBs that fail the initial CRC verification. The receiver first predicts the vehicle state represented in the corresponding BSM from previously recovered BSMs and characterizes the associated prediction uncertainty. The resulting predictive distribution is converted into bit-level soft priors for the temporally predictable BSM fields. These priors are mapped to the corresponding systematic variables of the LDPC decoder and combined with the original channel information for a second LDPC decoding pass. 
The following subsections describe the probabilistic vehicle-state prediction, prediction-derived bit-prior construction, analysis of probabilistic bit priors, and prediction-aided re-decoding procedures, respectively.

\subsection{Probabilistic Vehicle-State Prediction}

 \begin{figure} [t]
    \centering
    \includegraphics[width=0.97\linewidth]{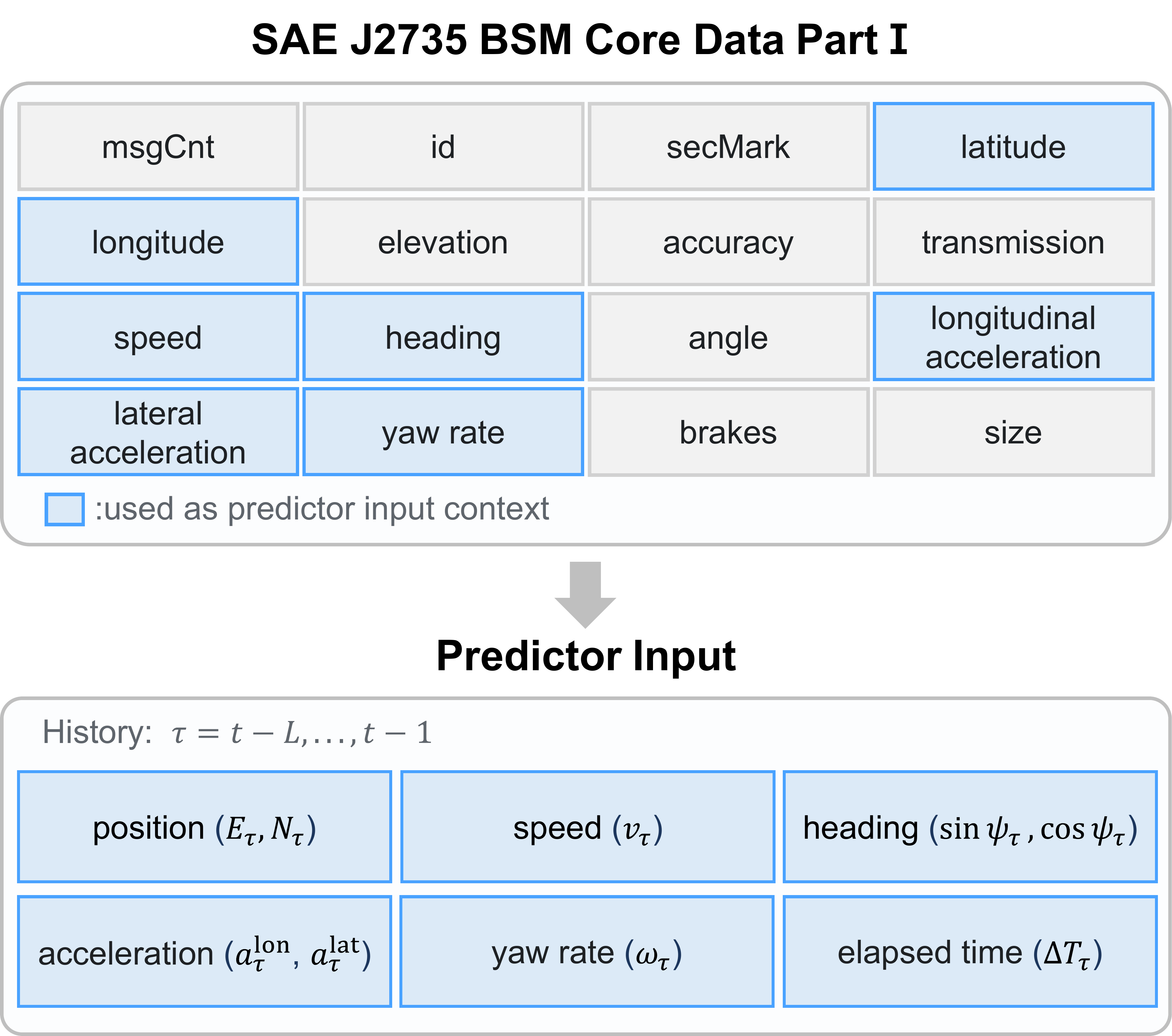}
    \caption{SAE J2735 BSM field structure and predictor input configuration for motion-state prediction.}
    \label{fig:data_structure}
\end{figure}

A single predicted position or speed value does not provide reliable bit-level confidence. Even minor prediction errors may lead to different serialized bit patterns. We use a predictive distribution to derive soft information for individual bits.
To obtain this distribution, a probabilistic gated recurrent unit (GRU) predictor~\cite{cho2014learning} models the target motion state from previously available BSMs. The predictor estimates the conditional distribution of the target state, capturing both the predicted state and its uncertainty. This distribution is used to derive soft information for the predictable bits.
For each history message at time $\tau$, the predictor input is represented by
\begin{equation}
\mathbf{x}_{\tau} = \left[E_{\tau}, N_{\tau}, v_{\tau},
\sin\psi_{\tau}, \cos\psi_{\tau}, a_{\tau}^{\mathrm{lon}}, a_{\tau}^{\mathrm{lat}},
{\omega}_{\tau}, \Delta T_{\tau} \right]^{\mathsf T} \in \mathbb{R}^{9},
\label{eq:gru_input}
\end{equation}
where $E_{\tau}$ and $N_{\tau}$ denote the vehicle position in a local East--North coordinate system, $v_{\tau}$ is the vehicle speed, $\psi_{\tau}$ is the heading angle, $a_{\tau}^{\mathrm{lon}}$ and $a_{\tau}^{\mathrm{lat}}$ denote the longitudinal and lateral accelerations, respectively, $\omega_{\tau}$ is the yaw rate, and $\Delta T_{\tau}$ is the elapsed time from the preceding message. Fig.~\ref{fig:data_structure} illustrates the correspondence between these predictor variables and the associated SAE J2735 BSM fields. In particular, the local position $(E_{\tau},N_{\tau})$ is derived from the latitude and longitude fields, while the heading is represented by its sine and cosine components.

The message history $\mathcal{H}_t$ is represented for prediction by the feature sequence
\begin{equation}
\mathbf{X}_{t} = \left[\mathbf{x}_{t-L},\ldots,\mathbf{x}_{t-1}\right]
\in \mathbb{R}^{L\times 9},
\label{eq:gru_history}
\end{equation}
from which the predictor models the conditional distribution of the target motion state
\begin{equation}
\mathbf{z}_{t} = \left[\Delta E_t,\,\Delta N_t,\,v_t\right]^{\mathsf T}
\in \mathbb{R}^{3}.
\label{eq:prediction_target}
\end{equation}
The displacements $\Delta E_t$ and $\Delta N_t$ are defined relative to the vehicle position in the latest BSM of the input history.
The predictor parameterizes this three-dimensional distribution using six output parameters: three conditional means and three corresponding standard deviations.

Both the input features and prediction targets are standardized using statistics computed from the training set. Let $\boldsymbol{\mu}_{x}$ and $\boldsymbol{\sigma}_{x}$ denote the input statistics, and let $\boldsymbol{\mu}_{z}$ and $\boldsymbol{\sigma}_{z}$ denote the target statistics. The normalized variables are
\begin{equation}
\widetilde{\mathbf{x}}_{\tau}
=\left(\mathbf{x}_{\tau}-\boldsymbol{\mu}_{x}\right)
\oslash \boldsymbol{\sigma}_{x}, \;
\widetilde{\mathbf{z}}_{t} = \left( \mathbf{z}_{t}-\boldsymbol{\mu}_{z}\right)
\oslash \boldsymbol{\sigma}_{z},
\label{eq:normalization}
\end{equation}
where $\oslash$ denotes element-wise division.
The normalized history is processed by a single-layer GRU. Let $\mathbf{h}_{t}$ denote the hidden state at the final history step. Two separate output heads generate the conditional mean and standard deviation:
\begin{align}
\mathbf{h}_{t} &= f_{\mathrm{GRU}}
\left(\widetilde{\mathbf{x}}_{t-L},\dots,\widetilde{\mathbf{x}}_{t-1}\right),
\label{eq:gru_hidden}
\\
\widetilde{\boldsymbol{\mu}}_{t} &= \mathbf{W}_{\mu}\mathbf{h}_{t} + \mathbf{b}_{\mu},
\label{eq:mean_head}
\\
\widetilde{\boldsymbol{\sigma}}_{t} &=
\operatorname{Softplus} \left(\mathbf{W}_{\sigma}\mathbf{h}_{t} + \mathbf{b}_{\sigma} \right) + \epsilon_{\sigma},
\label{eq:scale_head}
\end{align}
where $\widetilde{\boldsymbol{\mu}}_{t}$ and $\widetilde{\boldsymbol{\sigma}}_{t}$ denote the predicted mean and standard-deviation vectors in the normalized state domain, respectively. The matrices $\mathbf{W}_{\mu}$ and $\mathbf{W}_{\sigma}$ and the bias vectors $\mathbf{b}_{\mu}$ and $\mathbf{b}_{\sigma}$ are learnable parameters of the mean and scale output heads.
The function $\operatorname{Softplus}(u)=\log(1+e^{u})$ is applied element-wise to ensure positive standard deviations, and $\epsilon_{\sigma}>0$ provides a small lower bound for numerical stability.
Table~\ref{tab:gru_architecture} summarizes the architecture of the probabilistic GRU predictor.

\begin{table}[ht]
\centering
\caption{Architecture of the Probabilistic GRU Predictor}
\label{tab:gru_architecture}
\renewcommand{\arraystretch}{1.05}
\setlength{\tabcolsep}{5pt}
\small
\begin{tabular}{lc}
\hline
\textbf{Parameter} & \textbf{Value} \\
\hline
Input sequence dimension & $10\times9$ \\
GRU layers & 1 \\
GRU hidden size & 64 \\
Mean head & $64\rightarrow3$ \\
Scale head & $64\rightarrow3$ \\
Output dimension & 6 \\
Scale activation & Softplus $+10^{-4}$ \\
\hline
\end{tabular}
\end{table}

The predictor parameterizes a diagonal Gaussian distribution in the normalized target domain as
\begin{equation}
p\!\left(\widetilde{\mathbf{z}}_t\mid\mathbf{X}_t\right)
= \mathcal{N}\left(\widetilde{\boldsymbol{\mu}}_t,\,\operatorname{diag}
\left(\widetilde{\boldsymbol{\sigma}}_t^{2}\right)\right).
\label{eq:normalized_distribution}
\end{equation}
The diagonal covariance model provides a state-dependent uncertainty for each predicted motion variable while keeping the predictive distribution compact.
The model is trained by minimizing the negative log-likelihood (NLL) of the predicted diagonal Gaussian. For a mini-batch containing $N_{\rm batch}$ samples, the implemented loss is
\begin{equation}
\mathcal{L}_{\mathrm{NLL}} = \frac{1}{3N_{\rm batch}}\sum_{i=1}^{N_{\rm batch}}\sum_{d=1}^{3}
\left[\log \widetilde{\sigma}_{i,d}+\frac{\left(\widetilde{z}_{i,d}-\widetilde{\mu}_{i,d}\right)^2}{2\widetilde{\sigma}_{i,d}^{2}}\right].
\label{eq:gru_nll}
\end{equation}
By predicting both the mean and the standard deviation, the model accounts for input-dependent prediction uncertainty instead of assuming a fixed residual variance. 

During inference, the predicted parameters are transformed back to the physical domain according to
\begin{align}
\boldsymbol{\mu}_{t} &= \boldsymbol{\mu}_{z}+\boldsymbol{\sigma}_{z}
\odot\widetilde{\boldsymbol{\mu}}_{t},
\label{eq:mean_denorm} \\
\boldsymbol{\sigma}_{t} &= \boldsymbol{\sigma}_{z} \odot \widetilde{\boldsymbol{\sigma}}_{t},
\label{eq:std_denorm}
\end{align}
where $\odot$ denotes element-wise multiplication. The resulting predictive
distribution is
\begin{equation}
\mathcal{P}_{t} = \mathcal{N} \left( \boldsymbol{\mu}_{t}, \operatorname{diag}
\left(\boldsymbol{\sigma}_{t}^{2}\right)\right).
\label{eq:physical_distribution}
\end{equation}

To propagate this uncertainty to the serialized BSM bits, we draw $S$ Monte Carlo realizations from $\mathcal{P}_{t}$
\begin{equation}
\mathbf{z}_{t}^{(s)} = \boldsymbol{\mu}_{t} +
\boldsymbol{\sigma}_{t}\odot \boldsymbol{\epsilon}^{(s)}, \;
\boldsymbol{\epsilon}^{(s)} \sim \mathcal{N}(\mathbf{0},\mathbf{I}_{3}),
\;s=1,\ldots,S.
\label{eq:state_sampling}
\end{equation}
Each realization represents a plausible motion state for the target BSM.

\subsection{Prediction-Derived Bit-Prior Construction}

The predictive distribution is defined in the continuous motion-state domain, whereas LDPC decoding operates on binary variables. Instead of forming a single deterministic bit pattern, we translate the predictive distribution into marginal probabilities for the predictable BSM bits. These probabilities are used to construct soft priors for LDPC re-decoding.

For the $s$-th realization in \eqref{eq:state_sampling}, the predicted East--North displacement is transformed into geodetic coordinates using the latest available BSM as the local reference. Let $\mathbf{o}_{t-1}$ denote the geodetic reference associated with this message. The resulting candidate state is
\begin{equation}
\mathbf{s}_{t}^{(s)} =
\left[\phi_{t}^{(s)},\lambda_{t}^{(s)},v_{t}^{(s)}\right]^{\mathsf T},
\end{equation}
where $\phi_{t}^{(s)}$ and $\lambda_{t}^{(s)}$ denote the candidate latitude and longitude, respectively, and $v_{t}^{(s)}$ denotes the sampled vehicle speed. The geodetic coordinates are obtained as
\begin{equation}
\left[\phi_{t}^{(s)},\lambda_{t}^{(s)}\right]
=f_{\mathrm{geo}}\left(\Delta E_{t}^{(s)},\Delta N_{t}^{(s)};\mathbf{o}_{t-1}\right),
\label{eq:enu_to_geo}
\end{equation}
where $\Delta E_{t}^{(s)}$ and $\Delta N_{t}^{(s)}$ are the sampled East and North displacements, $\mathbf{o}_{t-1}$ specifies the latitude and longitude of the local reference position, and $f_{\mathrm{geo}}(\cdot,\cdot;\cdot)$ denotes the transformation from local East--North coordinates to geodetic latitude and longitude.

Each candidate state is converted according to the same SAE J2735 field quantization and UPER representation used for the transmitted BSM. Let
\begin{equation}
b_{t,j}^{(s)} = g_{\mathrm{UPER},j} \left(\mathbf{s}_{t}^{(s)}\right),
\; j\in\mathcal{I}_{\mathrm{pred}},
\label{eq:uper_candidate_bit}
\end{equation}
where $g_{\mathrm{UPER},j}(\cdot)\in\{0,1\}$ denotes the $j$-th predictable bit obtained from the J2735 representation. Under the considered Core Data Part I profile, $\mathcal{I}_{\mathrm{pred}}$ contains the serialized latitude, longitude, and speed bits. Thus, the field quantization, valid ranges, and bit representation follow the same J2735 rules as those applied to the transmitted payload.

For each predictable bit position $j$, the $S$ sampled motion states produce $S$ binary realizations $\{b_{t,j}^{(s)}\}_{s=1}^{S}$. The probability that the corresponding BSM bit equals 1 is calculated as the ratio of samples in which that bit takes the value 1:
\begin{equation}
\pi_{t,j} \triangleq \Pr\left(m_{t,j}=1\mid\mathbf{X}_{t}\right)
\approx\frac{1}{S}\sum_{s=1}^{S}b_{t,j}^{(s)}, \;
j\in\mathcal{I}_{\mathrm{pred}}.
\label{eq:bit_probability}
\end{equation}
The resulting probability reflects the uncertainty of the motion-state prediction at each predictable BSM bit position.

To prevent finite Monte Carlo sampling from assigning absolute confidence
to a bit, the estimated probability is clipped as
\begin{equation}
\bar{\pi}_{t,j} = \operatorname{clip}
\left(\pi_{t,j},\epsilon_{p},1-\epsilon_{p}\right),
\label{eq:prob_clip}
\end{equation}
where $\epsilon_{p}$ is a small positive constant that sets the lower and upper
probability limits. The clipping function is defined as
\begin{equation}
\operatorname{clip}(x,l,u) = \min\left(\max(x,l),u\right).
\end{equation}
This operation prevents probabilities of exactly zero or one and keeps the resulting bit LLRs finite.
The resulting probability is converted to a prior LLR as
\begin{equation}
L_{\mathrm{pri},t}(j) = \log \frac{\bar{\pi}_{t,j}} {1-\bar{\pi}_{t,j}},
\; j\in\mathcal{I}_{\mathrm{pred}}.
\label{eq:prior_llr}
\end{equation}
A maximum prior magnitude $L_{\max}>0$ is imposed to limit the influence of highly confident prediction-derived information. The clipped prior is given by
\begin{equation}
\widetilde{L}_{\mathrm{pri},t}(j) =
\operatorname{clip}\left(L_{\mathrm{pri},t}(j), -L_{\max}, L_{\max}\right).
\label{eq:prior_llr_clip}
\end{equation}
This restriction prevents the prior from dominating the channel evidence during BP decoding.

The UPER-domain priors are mapped to the corresponding systematic variables of the LDPC mother code. Let $\varphi(j)$ denote the mapping from the serialized BSM position $j\in\mathcal{I}_{\mathrm{pred}}$ to its associated systematic variable-node position.  The decoder-domain prior is defined element-wise as
\begin{equation}
L_{\mathrm{pri},t}^{\mathrm{LDPC}}(n) =
\begin{cases}
\widetilde{L}_{\mathrm{pri},t}(j), & n=\varphi(j),\; j\in\mathcal{I}_{\mathrm{pred}}, \\
0, & \text{otherwise}.
\end{cases}
\label{eq:ldpc_prior_vector}
\end{equation}
Collecting these elements over all LDPC variable-node positions forms the decoder-domain prior vector $\mathbf{L}_{\mathrm{pri},t}^{\mathrm{LDPC}}$. It provides soft side information in the same variable-node domain used by the BP decoder. 

\begin{algorithm}[t]
\caption{Proposed Prediction-Aided LDPC Re-Decoding}
\label{alg:proposed_recovery}
\begin{algorithmic}[1]

\Require $\eta_t=0$, $\mathbf{L}_{\mathrm{ch},t}$, $\mathcal{H}_t$, $S$, $\alpha$, $\epsilon_p$, $L_{\max}$, $I_{\max}$

\State $\mathcal{P}_t \gets f_{\mathrm{pred}}(\mathcal{H}_t)$

\State Draw $\{\mathbf{z}_{t}^{(s)}\}_{s=1}^{S}$ from $\mathcal{P}_t$

\For{$s=1,\ldots,S$}
    \State $\mathbf{s}_{t}^{(s)} \gets \big[f_{\mathrm{geo}}(\Delta E_t^{(s)},\Delta N_t^{(s)};\mathbf{o}_{t-1}), v_t^{(s)} \big]$
    \ForAll{$j\in\mathcal{I}_{\mathrm{pred}}$}
        \State $b_{t,j}^{(s)}\gets g_{\mathrm{UPER},j}(\mathbf{s}_{t}^{(s)})$
    \EndFor
\EndFor

\ForAll{$j\in\mathcal{I}_{\mathrm{pred}}$}
    \State$\displaystyle \pi_{t,j} \gets \frac{1}{S}\sum_{s=1}^{S} b_{t,j}^{(s)}$
    \State $\bar{\pi}_{t,j} \gets \operatorname{clip}(\pi_{t,j},\epsilon_p,1-\epsilon_p)$
    \State $\displaystyle \widetilde{L}_{\mathrm{pri},t}(j) \gets \operatorname{clip}( \log\frac{\bar{\pi}_{t,j}}{1-\bar{\pi}_{t,j}},-L_{\max},L_{\max})$
\EndFor

\State $\mathbf{L}_{\mathrm{ch},t}^{\mathrm{LDPC}} \gets f_{\mathrm{DRM}}(\mathbf{L}_{\mathrm{ch},t})$
\State
$\mathbf{L}_{t}^{\mathrm{rec}} \gets \mathbf{L}_{\mathrm{ch},t}^{\mathrm{LDPC}} + \alpha\mathbf{L}_{\mathrm{pri},t}^{\mathrm{LDPC}}$
\State $\hat{\mathbf{b}}_{t}^{(1)} \gets \mathcal{D}_{\mathrm{BP}} (\mathbf{L}_{t}^{\mathrm{rec}};I_{\max})$
\State \Return $\hat{\mathbf{b}}_{t}^{(1)}$
\end{algorithmic}
\end{algorithm}

\subsection{Analysis of Probabilistic Bit Priors}
\label{sec:prob_prior_analysis}
We compare two prior constructions that differ in their treatment of prediction uncertainty. The hard point prior uses the bit values obtained from a single predicted state, while the probabilistic prior incorporates prediction uncertainty into the LLR magnitude. Let $B_{t,j}\in\{0,1\}$ denote the binary random variable corresponding to the $j$-th predictable BSM bit at time $t$. Its true conditional bit probability is defined as
\begin{equation}
q_{t,j} = \Pr(B_{t,j}=1\mid\mathcal{H}_t).
\label{eq:true_bit_probability}
\end{equation}
The true conditional probability $q_{t,j}$ is not directly observable for an individual realization. The Monte Carlo estimate $\pi_{t,j}$ is intended to approximate this conditional probability. We analyze the idealized case $\pi_{t,j}=q_{t,j}$ to characterize the behavior of the proposed probabilistic prior. The probabilistic prior under the imposed magnitude constraint is
\begin{equation}
L_{t,j}^{\mathrm{prob}} =
\operatorname{clip}\left(\log\frac{q_{t,j}}{1-q_{t,j}},-L_{\max},L_{\max}\right).
\label{eq:prob_prior_optimal}
\end{equation}

For the hard point prior, let $\hat B_{t,j}\in\{0,1\}$ denote the bit obtained by serializing a single predicted motion state. Using the common magnitude bound $L_{\max}$, the hard point prior is
\begin{equation}
L_{t,j}^{\mathrm{hard}} = (2\hat B_{t,j}-1)L_{\max}.
\label{eq:hard_prior}
\end{equation}
For an arbitrary prior LLR $L$, the corresponding bit-one probability is
\begin{equation}
r(L) = \frac{1}{1+\exp(-L)} \in(0,1).
\end{equation}
The expected binary NLL conditioned on $\mathcal{H}_t$ is
\begin{equation}
\mathcal{R}(r(L);q_{t,j}) = -q_{t,j}\log r(L)-(1-q_{t,j})\log\left(1-r(L)\right).
\label{eq:prior_expected_nll}
\end{equation}

Under this calibrated case, the probabilistic LLR in \eqref{eq:prob_prior_optimal} minimizes the expected binary NLL in \eqref{eq:prior_expected_nll} over all priors satisfying $|L|\le L_{\max}$. First consider the unconstrained case. The excess expected NLL of an arbitrary prior LLR $L$ relative to the calibrated probability $q_{t,j}$ is
\begin{align}
&\mathcal{R}(r(L);q_{t,j}) - \mathcal{R}(q_{t,j};q_{t,j}) \nonumber\\
& \qquad = q_{t,j}\log\frac{q_{t,j}}{r(L)} +
(1-q_{t,j})\log\frac{1-q_{t,j}}{1-r(L)} \ge 0.
\label{eq:prior_kl_regret}
\end{align}
This quantity is the KL divergence between Bernoulli distributions with parameters $q_{t,j}$ and $r(L)$, and is minimized to zero when $r(L)=q_{t,j}$.

Under the magnitude constraint $|L|\le L_{\max}$, the minimum is obtained by clipping the LLR derived from the calibrated bit probability as in \eqref{eq:prob_prior_optimal}. As the bit prediction becomes more uncertain, $q_{t,j}$ approaches $0.5$, and the probabilistic prior LLR approaches zero. In contrast, the hard prior retains the fixed magnitude $L_{\max}$ regardless of prediction uncertainty. Therefore, an incorrect hard decision can introduce a strong conflicting prior even when the underlying prediction is uncertain. A detailed derivation and a complementary variance analysis are provided in Appendix~\ref{app:prob_prior_analysis}.

\subsection{Prediction-Aided Re-Decoding}

The prediction-aided recovery stage is activated only when the initially decoded TB fails CRC verification, i.e., $\eta_t=0$. No retransmission or additional channel observation is required. Instead, the same received soft information is reused together with the prediction-derived bit priors constructed in the previous subsection, as summarized in Algorithm~\ref{alg:proposed_recovery}. Let
\begin{equation}
\mathbf{L}_{\mathrm{ch},t}^{\mathrm{LDPC}} =
f_{\mathrm{DRM}}\left(\mathbf{L}_{\mathrm{ch},t}\right)
\label{eq:deratematched_llr}
\end{equation}
denote the channel LLRs represented in the LDPC mother-code variable-node domain.
The prediction-derived prior vector $\mathbf{L}_{\mathrm{pri},t}^{\mathrm{LDPC}}$ in \eqref{eq:ldpc_prior_vector} is defined in the same variable-node domain. For a packet entering the recovery stage, the decoder input is constructed as
\begin{equation}
\mathbf{L}_{t}^{\mathrm{rec}}
= \mathbf{L}_{\mathrm{ch},t}^{\mathrm{LDPC}}
+ \alpha \mathbf{L}_{\mathrm{pri},t}^{\mathrm{LDPC}},
\label{eq:llr_combination}
\end{equation}
where $\alpha\geq0$ controls the contribution of the predictive prior.

A fresh BP decoding pass is then performed using $\mathbf{L}_{t}^{\mathrm{rec}}$, which is already expressed in the LDPC variable-node domain:
\begin{equation}
\hat{\mathbf{b}}_{t}^{(1)}
= \mathcal{D}_{\mathrm{BP}}
\left(\mathbf{L}_{t}^{\mathrm{rec}}; I_{\max}\right)
= \left[\hat{\mathbf{a}}_{t}^{(1)}, \hat{\mathbf{r}}_{t}^{(1)}\right],
\label{eq:second_bp}
\end{equation}
with
\begin{equation}
\hat{\mathbf{a}}_{t}^{(1)}
= \left[\hat{\mathbf{m}}_{t}^{(1)}, \hat{\mathbf{q}}_{t}^{(1)}\right].
\end{equation}
The superscript $(1)$ identifies estimates obtained from the prediction-aided decoding pass. The BP decoder is reinitialized using $\mathbf{L}_{t}^{\mathrm{rec}}$, without retaining messages from the initial decoding pass. With a fixed budget of $I_{\max}$ iterations for each BP pass, the proposed receiver performs $2I_{\max}$ iterations for a packet entering the recovery stage, while a packet that passes the initial CRC verification requires only $I_{\max}$ iterations.

\section{Simulation Results} \label{sec:results}

\subsection{Simulation Setup}
The simulations use the Advanced Messaging Concept Development provided by the U.S. Department of Transportation~\cite{USDOT_AMCD}. Continuous vehicle trajectories are constructed by sender and timestamp, and each prediction sample uses the previous $L=10$ BSMs to predict the motion state of the next BSM. The resulting prediction sequences are divided into 1,076,362 training, 114,171 validation, and 151,868 test samples. 
For the PHY evaluation, 200,000 TB transmissions are simulated at each $E_b/N_0$ point using the test data.

The AMCD logs provide vehicle application-state information rather than bit-exact transmitted BSM packets. We reconstruct each BSM using a controlled SAE J2735:2024 Core Data Part I profile and serialize it with ASN.1 UPER. The prediction-derived priors are applied to the latitude, longitude, and speed fields, which occupy 31, 32, and 13 bits, respectively, for a total of 76 predictable bits. The remaining BSM and link parameters are summarized in Table~\ref{tab:sim_config}.

\begin{table}[ht]
\centering
\caption{Standards-Based BSM and NR Link Configuration}
\label{tab:sim_config}
\renewcommand{\arraystretch}{1.05}
\setlength{\tabcolsep}{5pt}
\small
\begin{tabular}{ll}
\hline
\textbf{Parameter} & \textbf{Setting} \\
\hline
BSM specification & SAE J2735 Core Data Part I \\
Serialization & ASN.1 UPER \\
Serialized BSM length $M$ & 320 bits \\
Resource-related parameters & TS 38.533 Rel.~17, CD.1A HD \\
SL-SCH channel coding & TS 38.212 Rel.~17 \\
Reference allocation & 10 RB, 10 PSSCH symbols/slot \\
TB size $A$ & 672 bits \\
TB CRC & CRC16 \\
CRC-appended TB size $B$ & 688 bits \\
NR LDPC base graph & BG2 \\
Lifting size $Z_c$ & 72 \\
LDPC systematic dimension $K$ & 720 bits \\
Filler bits $F$ & 32 bits \\ 
Rate-matched length $E$ & 2160 bits \\
Modulation & QPSK \\
\hline
\end{tabular}
\end{table}

The link-level simulations are conducted over an AWGN channel. 
The resource-related parameters are derived from the CD.1A HD PSSCH reference measurement channel in 3GPP TS~38.533 Rel.~17~\cite{ts38533rel17}, while SL-SCH channel coding, including TB CRC attachment, LDPC encoding, and rate matching, follows 3GPP TS~38.212 Rel.~17~\cite{ts38212rel17}. The LDPC decoding procedure is implemented using Sionna~\cite{hoydis2022sionna}. The $E_b/N_0$ range is varied from 0 to 1~dB in 0.125~dB increments. The maximum number of BP iterations is set to $I_{\max}=40$. The initial decoder uses this iteration limit. If the CRC verification fails, BP decoding is restarted with the same limit in the prediction-aided recovery stage.

\begin{figure}[ht]
    \centering
    \includegraphics[width=0.85\linewidth]{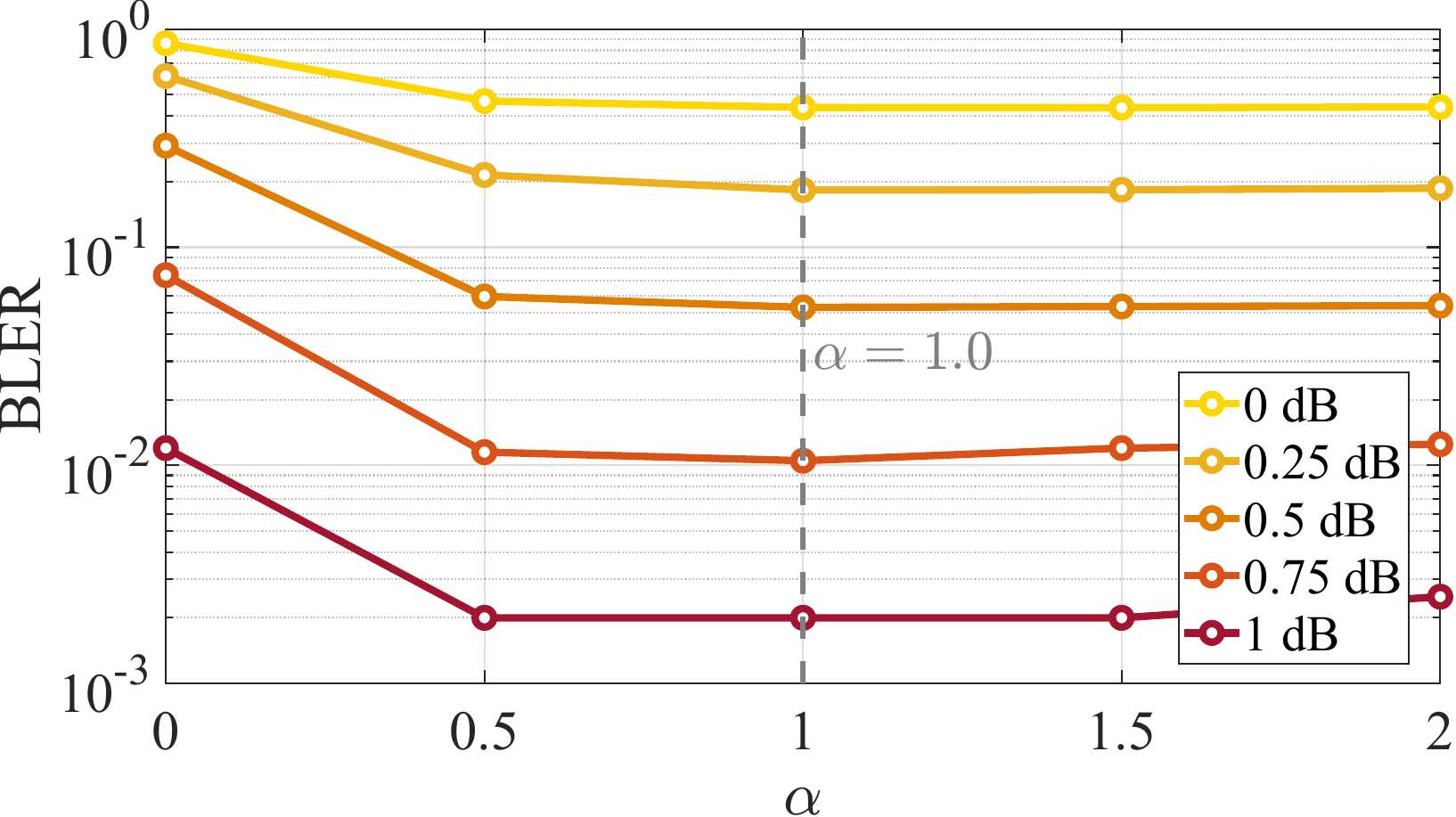}
    \caption{Sensitivity of prediction-aided decoding performance to the
    prior scaling factor $\alpha$ on the validation set.}
    \label{fig:alpha_sensitivity}
\end{figure}

For prediction-aided recovery, $S=512$ Monte Carlo samples are drawn from the predictive motion-state distribution for each target BSM. The empirical bit probabilities are clipped with $\epsilon_p=10^{-3}$, and the prior LLR magnitude is limited to $L_{\max}=6$.
Fig.~\ref{fig:alpha_sensitivity} shows the BLER sensitivity to $\alpha$ at representative $E_b/N_0$ values. The validation results show a stable operating region around $\alpha=1$. Therefore, the prior scaling factor is set to $\alpha=1$ based on the validation set.

\subsection{Baselines and Evaluation Metrics}

The proposed receiver is compared with conventional LDPC decoding and
prediction-based recovery baselines under the same payload and channel
realizations. The considered methods are summarized as follows:
\begin{itemize}
    \item \textbf{BP-40}: Conventional LDPC decoding with a maximum of
    40 BP iterations and no prediction-derived prior.

    \item \textbf{BP-80}: Conventional LDPC decoding with a maximum of
    80 BP iterations. 
    
    \item \textbf{CV}: Prediction-aided recovery using a constant-velocity motion predictor with the same probabilistic bit-prior construction and LDPC re-decoding procedure as the proposed method. Its predictive uncertainty is modeled using validation-set residual standard deviations.

    \item \textbf{Kalman}: Prediction-aided recovery using a Kalman-based motion predictor with the same probabilistic bit-prior construction and LDPC re-decoding procedure as the proposed method. Its predictive uncertainty is also obtained from validation-set residual standard deviations.

    \item \textbf{Proposed GRU}: Prediction-aided recovery using the proposed probabilistic GRU predictor. The predicted motion-state distribution is converted into uncertainty-aware bit priors using the UPER-based prior construction.

    \item \textbf{Hard Point Prior}: Prediction-aided recovery using a serialized point prediction with fixed-magnitude prior LLRs of $L_{\max}$, without uncertainty-dependent scaling.

    \item \textbf{Field Oracle}: The true values of the predictable BSM bits are used as oracle priors, with $+L_{\max}$ assigned to a transmitted bit of 1 and $-L_{\max}$ to a transmitted bit of 0. The priors are injected only into the corresponding systematic LDPC variables. This method serves as an oracle reference under the adopted prior magnitude and predictable-field setting and is not deployable.
\end{itemize}

Performance is evaluated using the following metrics:

\begin{itemize}
    \item \textbf{Block error rate (BLER):}
    BLER is defined as the fraction of transmitted TBs that are incorrectly decoded after the complete receiver procedure.

    \item \textbf{CRC Recovery Rate:}
    This metric measures the fraction of TBs that fail the initial CRC verification but are successfully recovered by the subsequent decoding stage.

    \item \textbf{Correct Sign:}
    This metric measures the fraction of predictable BSM bits for which the prior LLR has the correct sign. A larger value indicates more accurate bit-direction prediction.

    \item \textbf{Bit NLL:}
    Bit NLL evaluates the probability assigned by the prior to the transmitted bit value. A smaller value indicates that the predicted bit probabilities are both accurate and appropriately confident. 

    \item \textbf{Brier score:}
    The Brier score measures the mean squared error between the predicted probability of bit 1 and the transmitted binary bit value. A smaller value indicates better probabilistic prediction and provides a complementary measure of probabilistic prediction quality.

    \item \textbf{Bit error rate (BER):}
    BER is used to examine the bit-level effect of prediction-derived priors.
\end{itemize}

\subsection{Results}

\subsubsection{Overall BLER and CRC Recovery Performance}

\begin{figure}[t]
    \centering
    \includegraphics[width=0.85\linewidth]{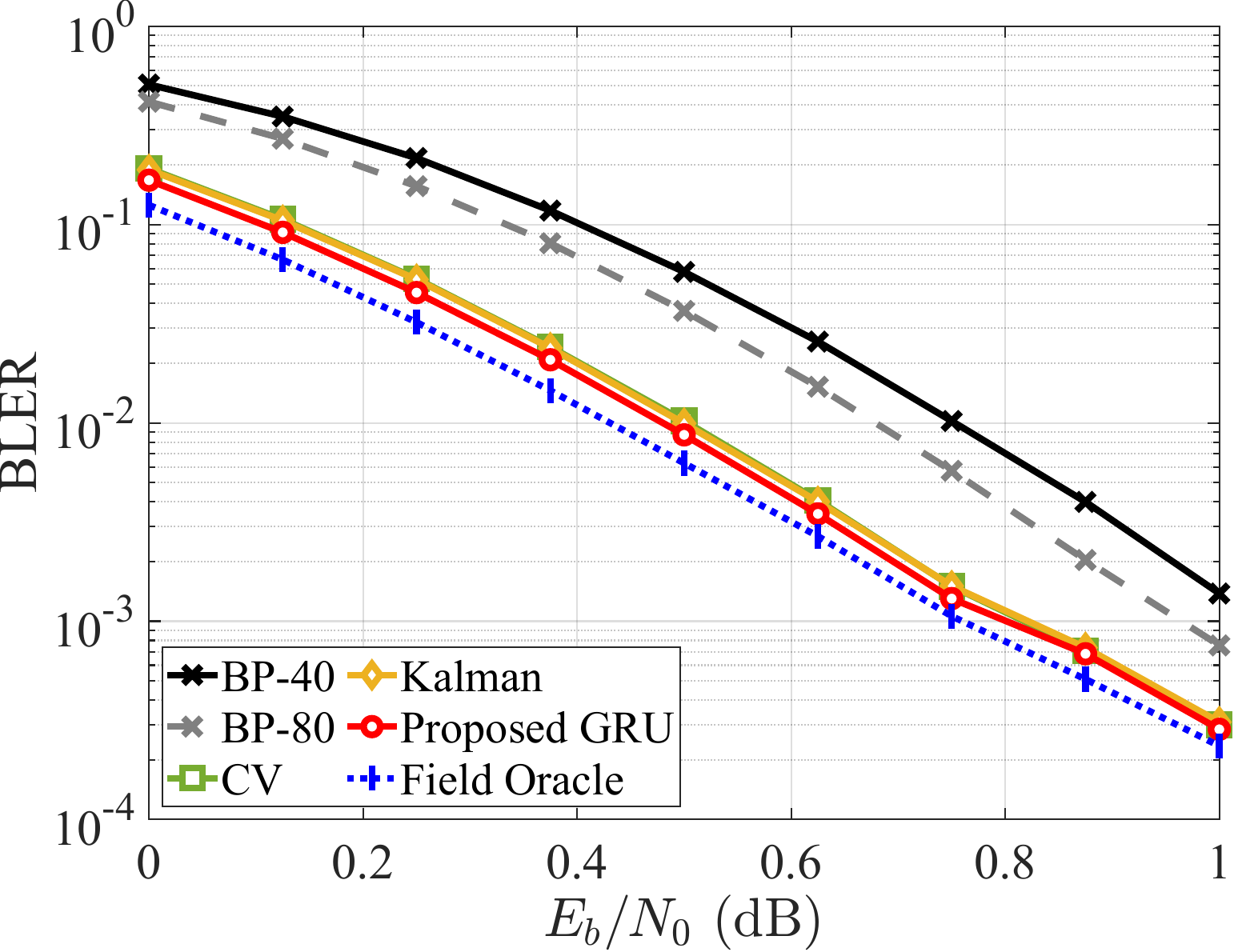}
    \caption{Transport-block BLER performance versus $E_b/N_0$ for
    conventional BP decoding and prediction-aided recovery methods.}
    \label{fig:main_bler}
\end{figure}

Fig.~\ref{fig:main_bler} compares the BLER performance of the conventional BP baselines and the prediction-aided recovery methods. Across the entire evaluated $E_b/N_0$ range, all prediction-aided methods achieve lower BLER than conventional BP decoding, indicating that the motion-derived priors provide useful side information beyond additional BP iterations alone. The proposed GRU consistently gives the best performance among the deployable prediction methods, while the Field Oracle provides the lowest BLER as an upper-bound reference. 

\begin{figure}[t]
    \centering
    \includegraphics[width=0.85\linewidth]{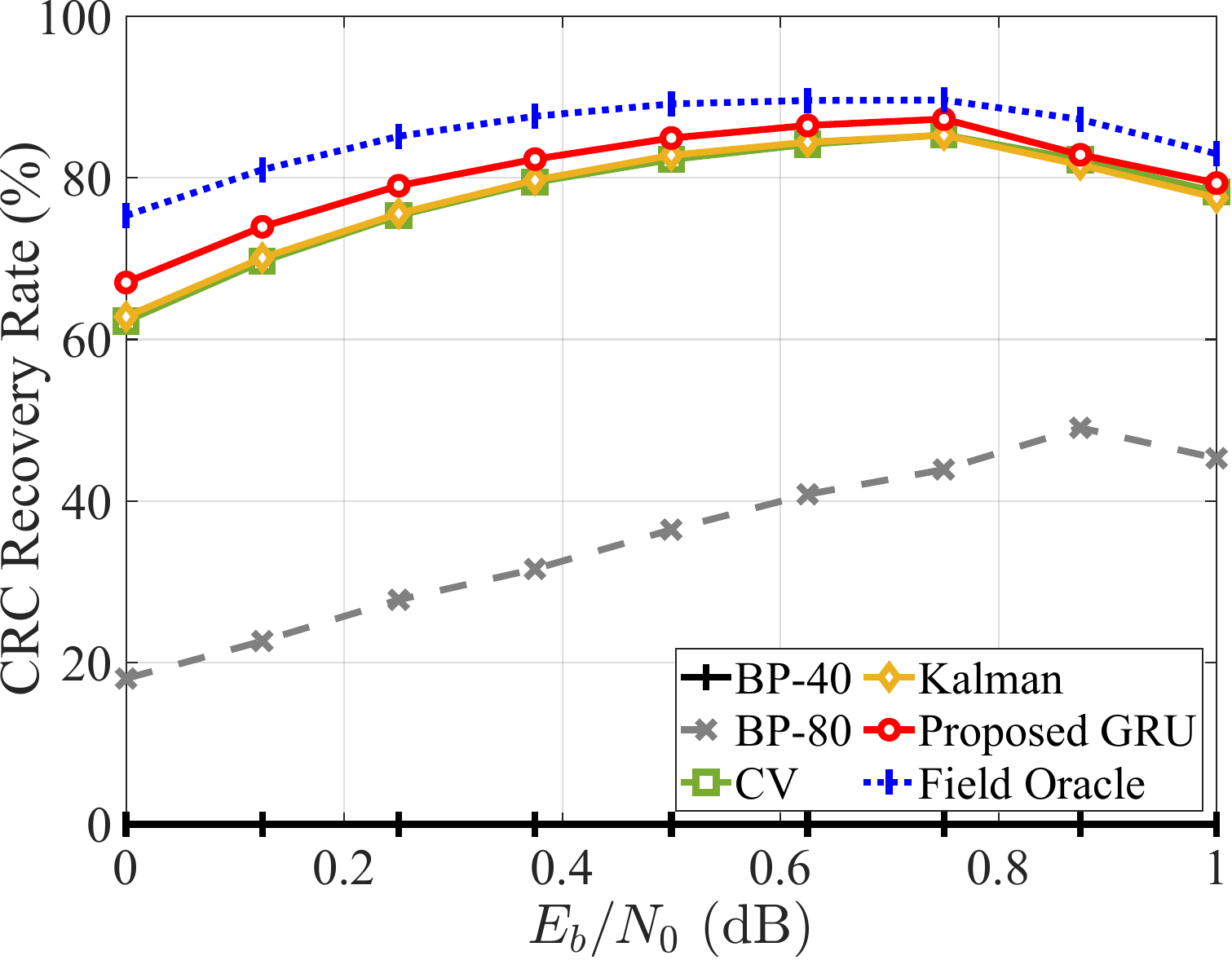}
    \caption{CRC Recovery Rate among TBs that fail the initial decoding pass.}
    \label{fig:main_recovery}
\end{figure}

\begin{figure*}[t]
    \centering

    \subfloat[CV]{
        \includegraphics[width=0.3\textwidth]
        {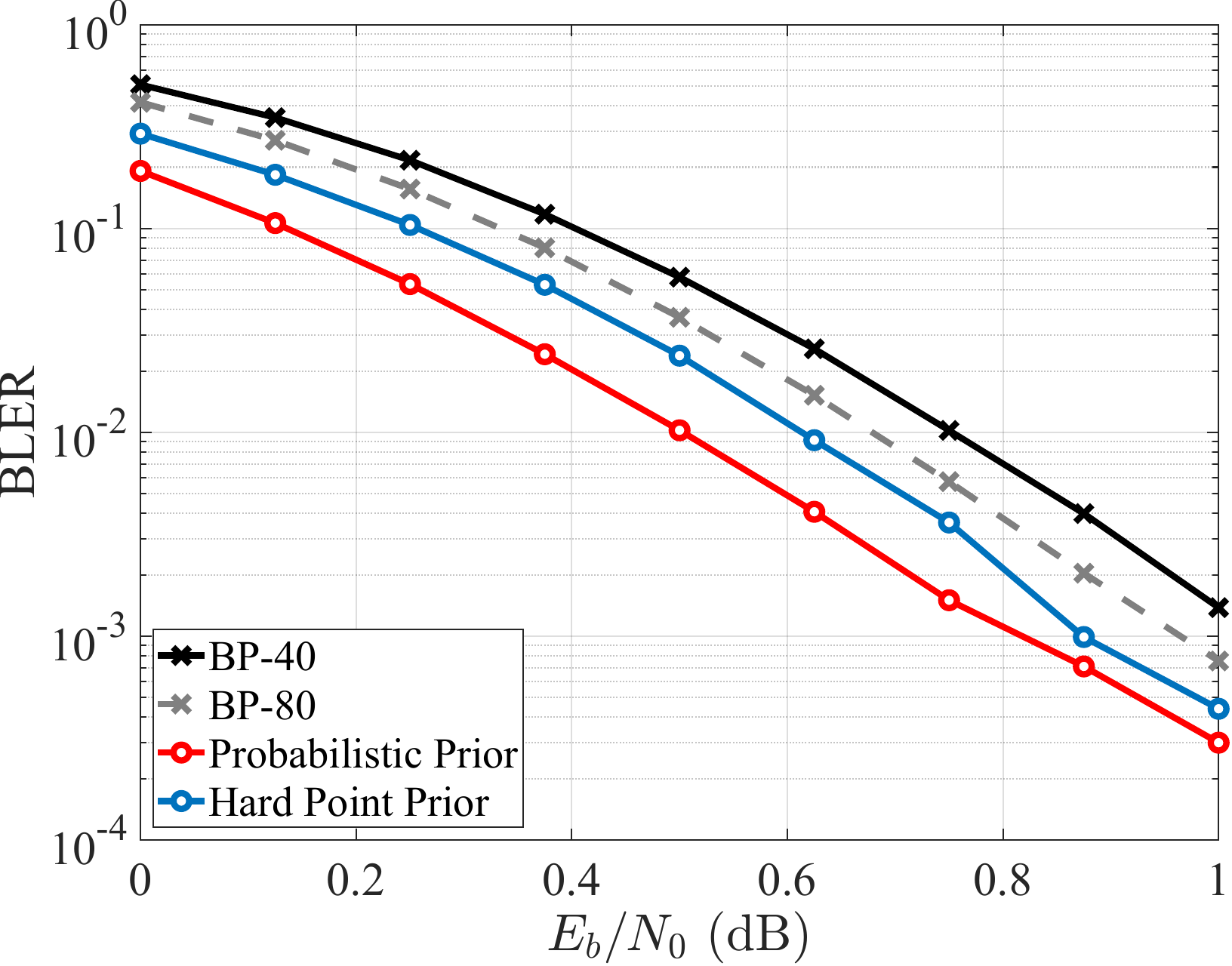}
        \label{fig:prob_hard_cv}
    }
    \hfill
    \subfloat[Kalman]{
        \includegraphics[width=0.3\textwidth]
        {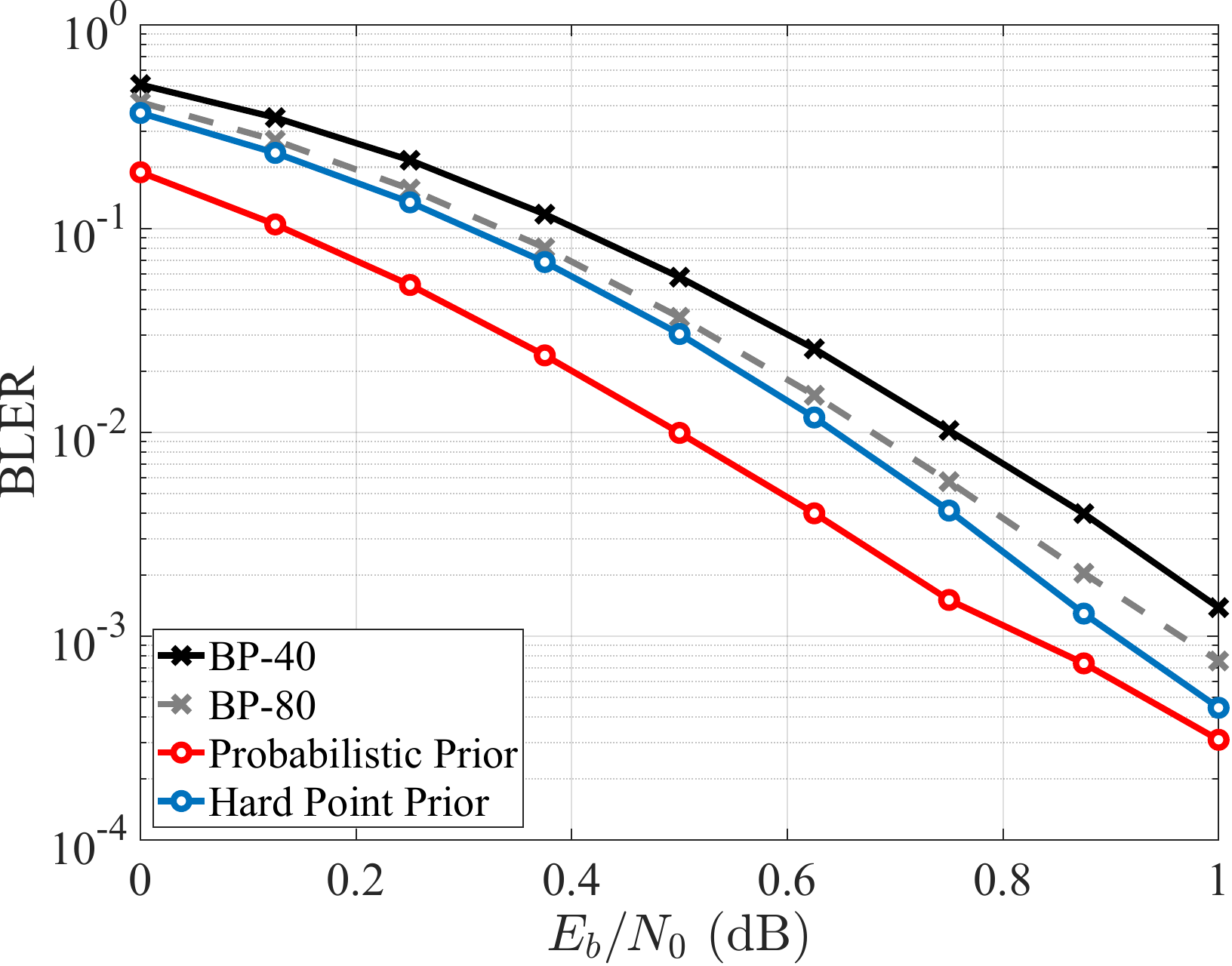}
        \label{fig:prob_hard_kalman}
    }
    \hfill
    \subfloat[Proposed GRU]{
        \includegraphics[width=0.3\textwidth]
        {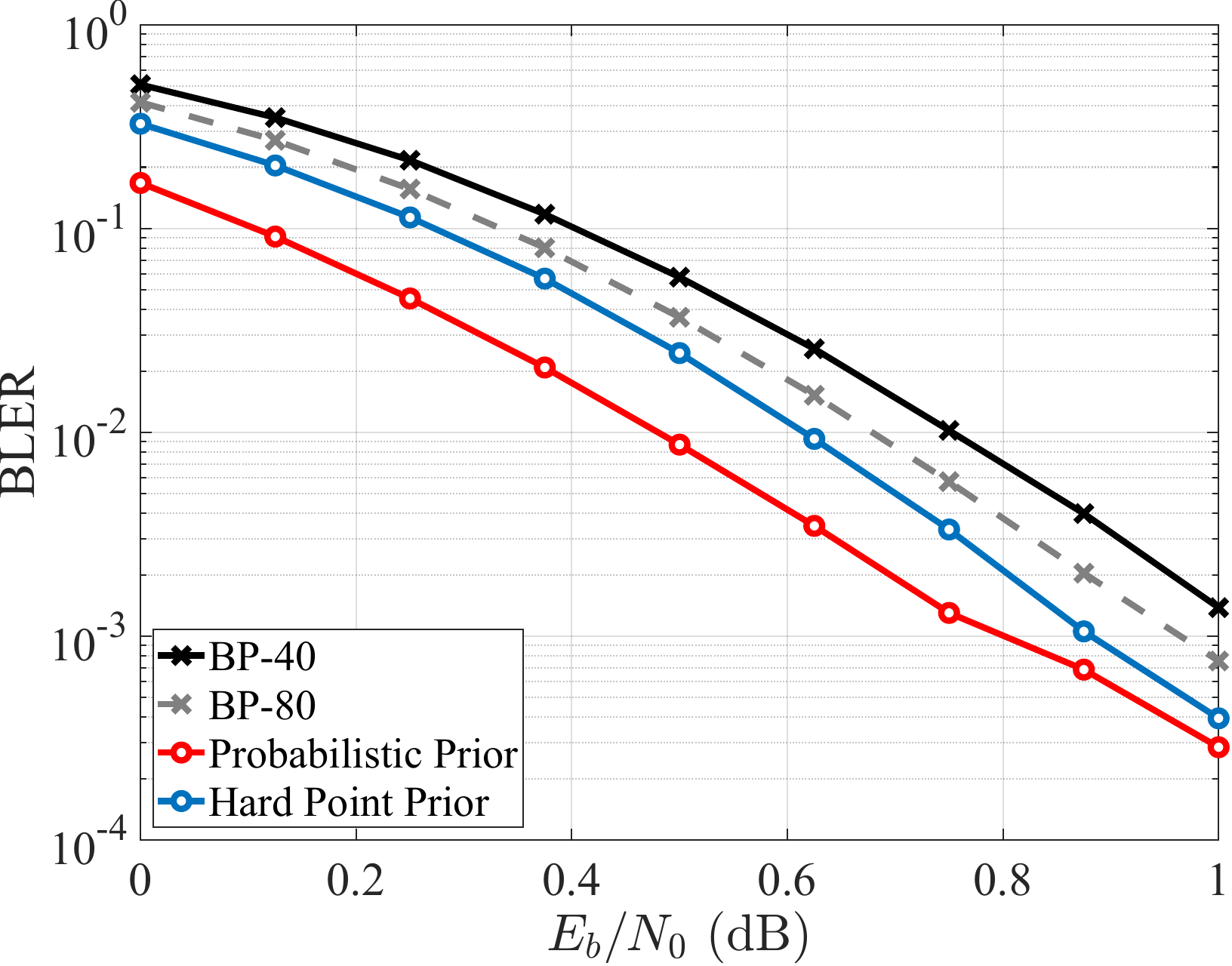}
        \label{fig:prob_hard_gru}
    }

    \caption{BLER comparison between the probabilistic and hard point priors for different motion predictors.}
    \label{fig:prob_vs_hard}
\end{figure*}

\begin{figure*}[t]
    \centering

    \subfloat[Correct Sign]{
        \includegraphics[width=0.3\textwidth]{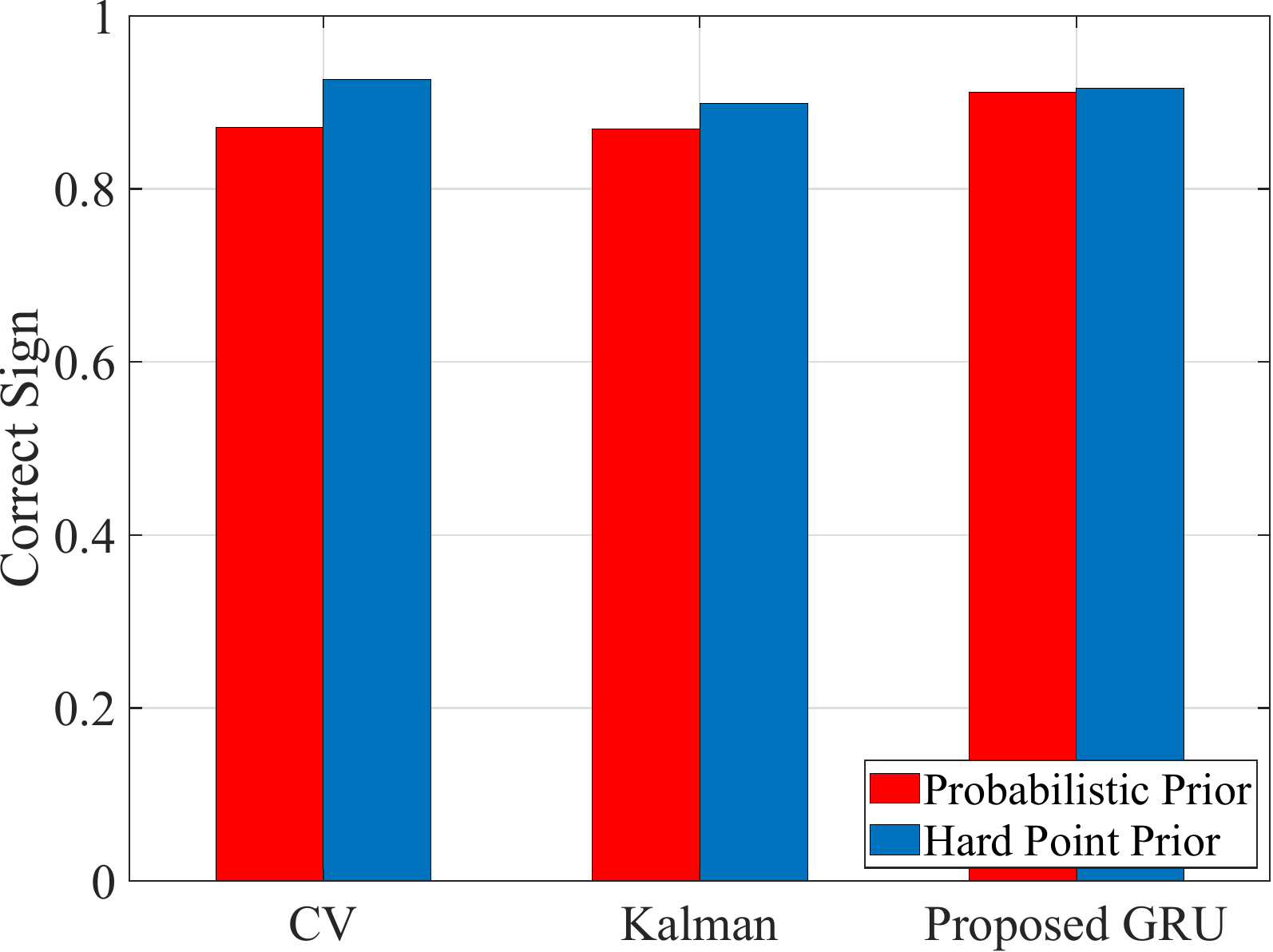}
        \label{fig:prior_quality_sign}}
    \hfill
    \subfloat[Bit NLL]{
        \includegraphics[width=0.3\textwidth]{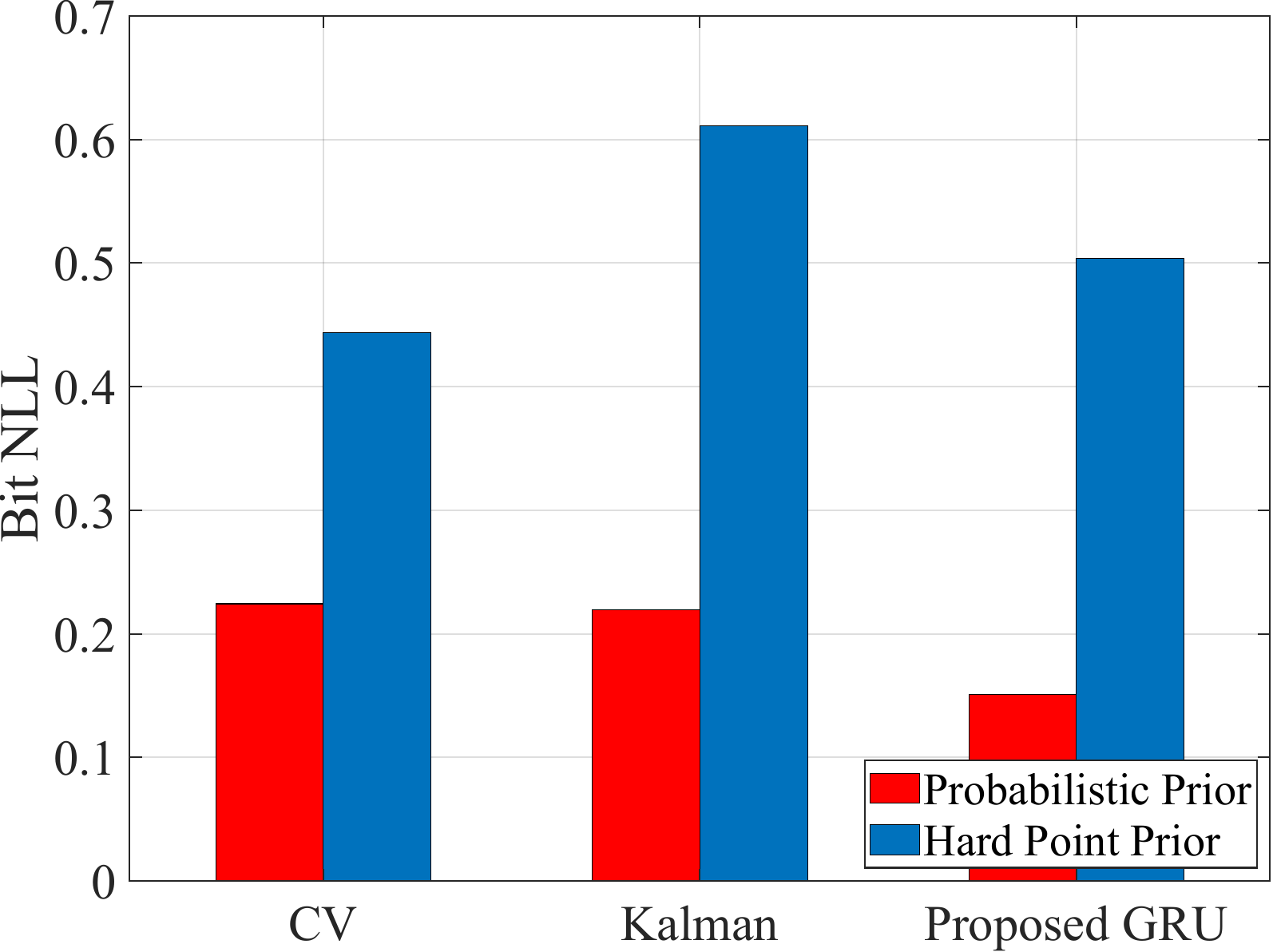}
        \label{fig:prior_quality_nll}}
    \hfill
    \subfloat[Brier score]{
        \includegraphics[width=0.3\textwidth]{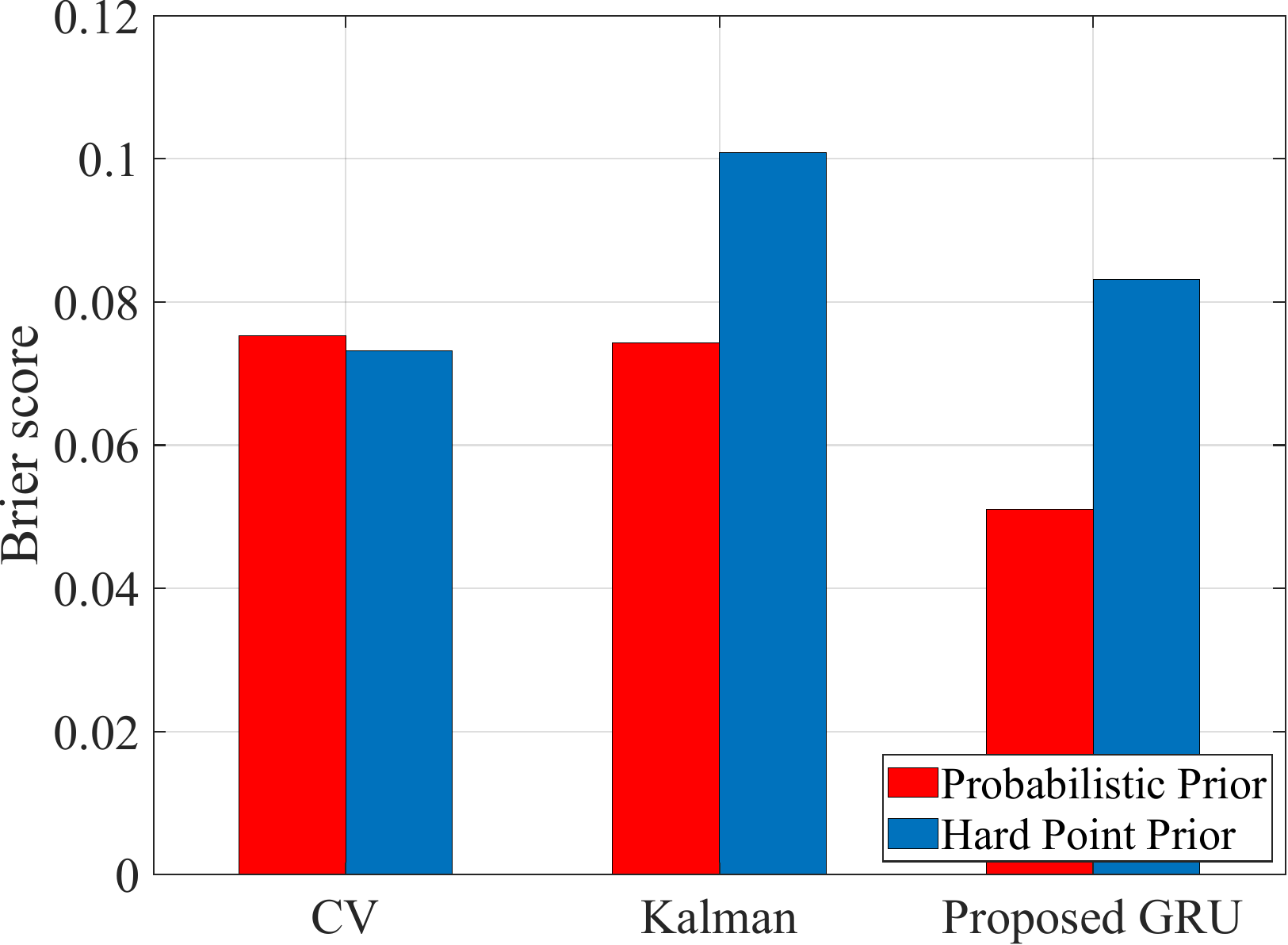}
        \label{fig:prior_quality_brier}}

    \caption{Bit-prior quality comparison between the probabilistic and hard point priors for different motion predictors.}
    \label{fig:prior_quality_prob_hard}
\end{figure*}

Fig.~\ref{fig:main_recovery} further evaluates the fraction of TBs recovered after an initial BP-40 CRC failure. The prediction-aided methods recover a substantially larger portion of the failed TBs than BP-80 across the considered $E_b/N_0$ range. At $E_b/N_0=0.75$~dB, the proposed GRU recovers 87.27\% of the initially failed TBs, compared with 43.91\% for BP-80. These results show that probabilistic motion-derived priors provide additional side information for recovery that cannot be obtained by simply increasing the BP iteration budget.

\subsubsection{Effect of Probabilistic Bit Priors}

Fig.~\ref{fig:prob_vs_hard} compares the probabilistic prior with the hard point prior for the same motion predictors. The probabilistic prior converts the predicted bit probability into an LLR whose magnitude reflects prediction confidence, whereas the hard point prior assigns a fixed magnitude $L_{\max}$ according to a single predicted bit value. The probabilistic prior consistently achieves lower BLER for CV, Kalman, and the proposed GRU. Since the underlying predictor and re-decoding procedure are unchanged, this comparison isolates the effect of prior construction. The result indicates that preserving prediction uncertainty in the LLR magnitude is more effective for decoding than converting the same prediction into a hard point prior.

Fig.~\ref{fig:prior_quality_prob_hard} explains this behavior. The hard point prior achieves a higher Correct Sign for all three predictors, indicating that a single point estimate selects the transmitted bit direction more frequently. However, this higher sign accuracy does not lead to lower BLER in Fig.~\ref{fig:prob_vs_hard}.
The Field Oracle also uses the fixed magnitude $L_{\max}$, but always assigns the correct prior direction and achieves the best recovery performance in Figs.~\ref{fig:main_bler} and~\ref{fig:main_recovery}. This result shows that a large prior magnitude is not itself harmful. The problem arises when the same large magnitude is assigned to uncertain or incorrect predictions, which can strongly conflict with the channel evidence. Even if such errors occur less frequently, a wrong-sign prior with large magnitude may dominate the channel evidence and increase decoding errors. The probabilistic prior mitigates this effect by reducing the LLR magnitude for uncertain predictions.
In contrast, the probabilistic prior achieves a substantially lower bit NLL for all predictors. This result is consistent with the analysis in Section~\ref{sec:prob_prior_analysis}, where the probabilistic prior minimizes the expected bit NLL under the calibrated model. The lower NLL shows that the probabilistic prior assigns confidence more consistently with the uncertainty of the predicted bit.
The Brier score shows a similar trend for Kalman and the proposed GRU, while CV is a small exception, for which the hard point prior achieves a slightly lower score. Its higher Correct Sign allows the hard prior to benefit from near-certain probabilities on many correctly predicted bits. However, the same confidence is assigned to incorrect predictions, which are penalized more strongly by bit NLL and can adversely affect LDPC decoding. Therefore, the NLL and BLER results show that reliable decoder priors require confidence to reflect prediction uncertainty, not only accurate bit direction.

\subsubsection{Propagation of Prediction-Derived Priors}

\begin{figure*}[t]
    \centering

    \subfloat[CV]{
        \includegraphics[width=0.3\textwidth]
        {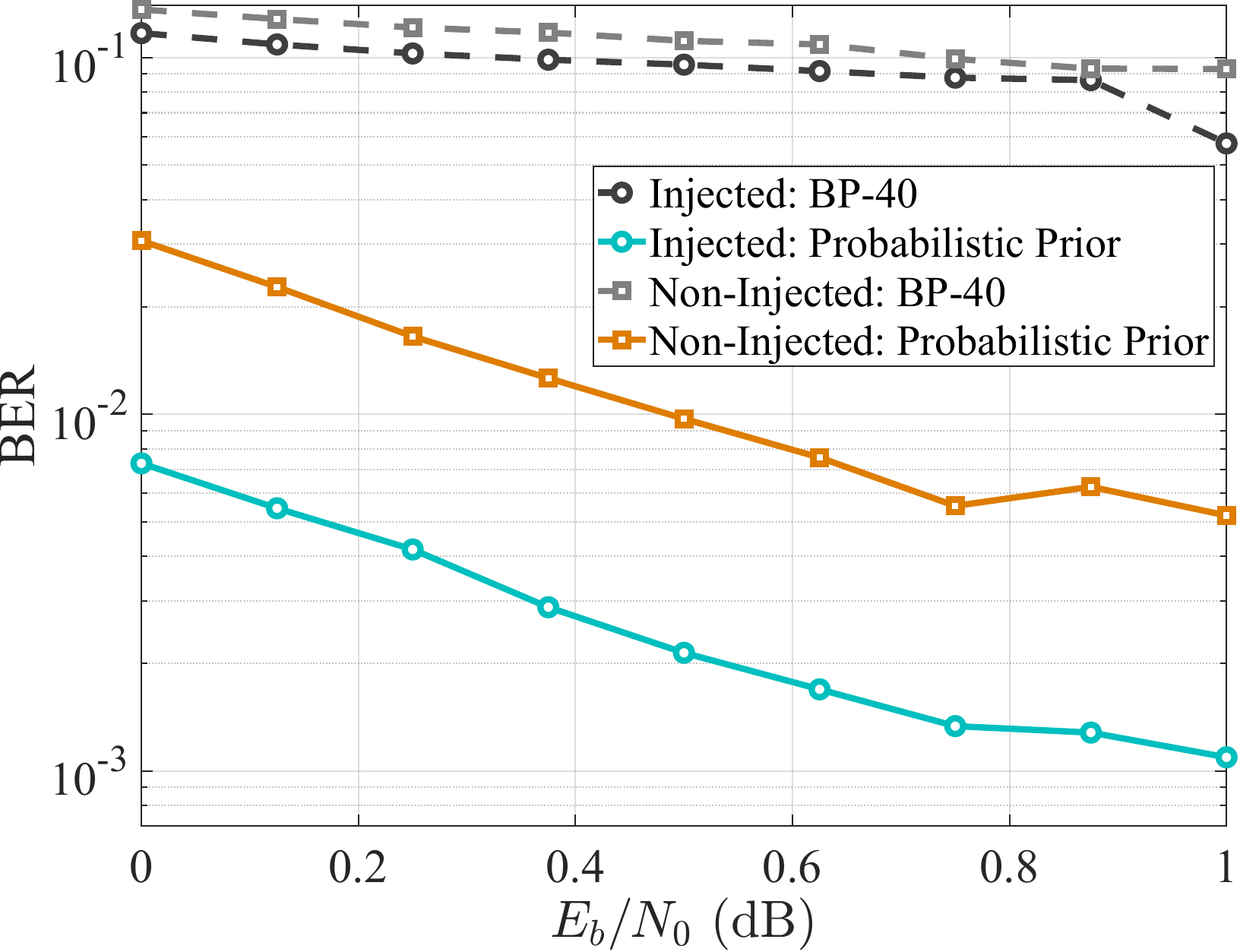}
        \label{fig:injection_cv}
    }
    \hfill
    \subfloat[Kalman]{
        \includegraphics[width=0.3\textwidth]
        {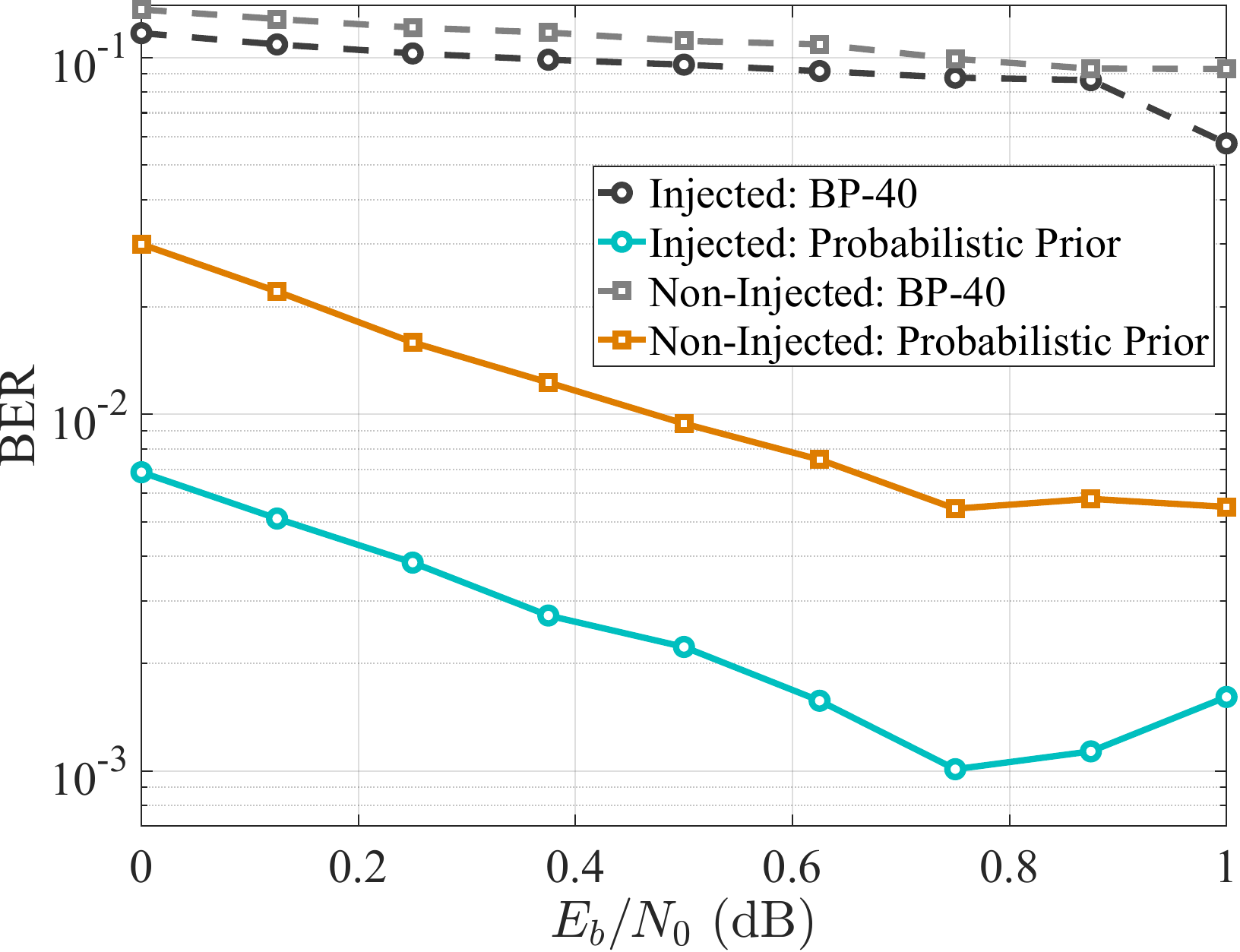}
        \label{fig:injection_kalman}
    }
    \hfill
    \subfloat[Proposed GRU]{
        \includegraphics[width=0.3\textwidth]
        {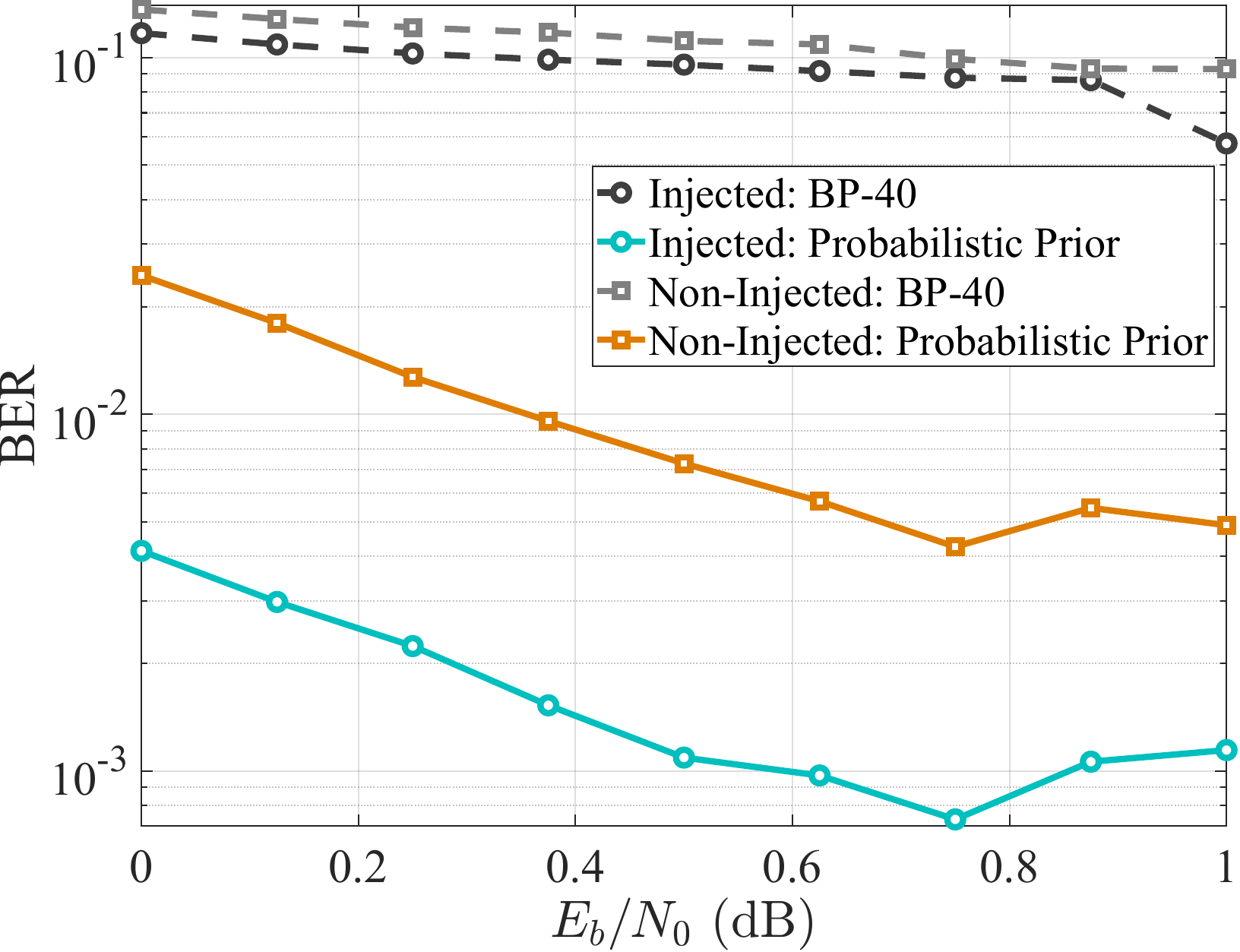}
        \label{fig:injection_gru}
    }

    \caption{BER on directly prior-injected and non-injected TB bits for different motion predictors, evaluated over TBs that fail the initial BP-40 CRC verification.}
    \label{fig:injected_noninjected}
\end{figure*}

Fig.~\ref{fig:injected_noninjected} examines the bit-level effect of the prediction-derived priors on TBs that fail the initial CRC verification. The conditional BER is evaluated separately for the 76 predictable BSM bits that directly receive the prior and the remaining 596 bits that receive no direct prediction-derived information.
For the directly injected bits, all three prediction methods substantially reduce BER over the considered $E_b/N_0$ range. This behavior reflects the direct contribution of the prediction-derived soft information to the corresponding systematic LDPC variables. The proposed GRU generally provides the lowest BER, consistent with its higher-quality priors.
A similar reduction is also observed for the non-injected bits. Their improvement indicates that the effect of the injected soft information propagates through the LDPC Tanner graph during BP re-decoding. The check-node and variable-node updates allow the updated beliefs at the prior-assisted variables to influence other coded variables through the parity constraints.

\subsubsection{Implementation Latency}

Table~\ref{tab:latency} reports the batch-1 software latency of the proposed receiver on an NVIDIA RTX 4090. The BP decoder is executed in compiled mode, and CRC verification is included in the decoding latency. For a message entering the recovery stage, the proposed end-to-end path includes the initial BP-40 decoding, GRU inference, probabilistic prior construction with 512 Monte Carlo samples, and the second prior-assisted BP-40 decoding pass. The reported runtime results are used to quantify the computational overhead introduced by the prediction-aided recovery stage relative to conventional LDPC decoding under the same software implementation.

\begin{table}[ht]
\centering
\caption{Batch-1 software decoding latency at $E_b/N_0=0.5$~dB on an NVIDIA RTX 4090.}
\label{tab:latency}
\begin{tabular}{lc}
\hline
\textbf{Decoding path} & \textbf{Mean (ms)} \\
\hline
BP-40 + CRC & 14.99  \\
BP-80 + CRC & 29.65  \\
Proposed recovery branch & 30.54 \\
CRC-gated proposed average & 15.88 \\
\hline
\end{tabular}
\end{table}

The failure-conditioned proposed recovery path requires 30.54~ms, which is close to the 29.65~ms latency of BP-80. This result indicates that the GRU inference and probabilistic prior construction introduce limited additional overhead relative to the two BP-40 decoding passes. However, the recovery stage is activated only when the initial BP-40 decoding fails CRC verification. At $E_b/N_0=0.5$~dB, the initial failure rate is 5.77\%, resulting in an average software decoding latency of 15.88~ms. This reduction arises from applying the prediction-aided recovery stage only to failed messages.

Overall, the simulation results demonstrate that prediction-derived probabilistic priors provide effective soft side information for LDPC recovery. Their benefit is observed across different motion predictors and is greater than that obtained by simply increasing the BP iteration budget. The results also show that preserving prediction uncertainty in the prior magnitude is important for effective recovery. The BER reduction on both directly injected and non-injected bits further shows that the resulting information propagates through the LDPC graph and contributes to recovery beyond the predicted bits.

\section{Conclusion} \label{sec:conclusion}

This paper proposed an uncertainty-aware prediction-aided LDPC recovery framework for SAE J2735 V2X safety messages. The proposed method exploits the temporal correlation of consecutive BSMs to assist the recovery of TBs that fail the initial CRC verification. Rather than relying on a deterministic point prediction, the receiver represents the predicted vehicle state as a distribution and converts the associated uncertainty into probabilistic bit-level prior LLRs. The resulting LLR magnitudes reflect prediction reliability and provide soft side information for a second LDPC decoding pass. The results show that preserving prediction uncertainty provides more effective decoder side information than the hard point prior across different motion predictors. The improvement observed on non-injected bits further indicates that the prediction-derived information propagates through the LDPC graph beyond the directly predicted BSM fields. The proposed recovery procedure operates only at the receiver and preserves the SAE J2735 message representation and NR SL-SCH channel-coding procedure without requiring retransmission or additional signaling. This design preserves standardized transmission procedures while enabling application-level temporal information to assist channel decoding.

\appendices
\section{Derivation of the Probabilistic-Prior Analysis}
\label{app:prob_prior_analysis}

\subsection{Expected-NLL Optimality}

Using the definition of $r(L)$, the expected binary NLL in \eqref{eq:prior_expected_nll} can be written directly as a function of $L$:
\begin{align}
\mathcal{R}(r(L);q_{t,j}) &= -q_{t,j}\log r(L) -(1-q_{t,j})\log(1-r(L)) \nonumber\\
&= \log\left(1+\exp(L)\right)-q_{t,j}L.
\label{eq:app_nll}
\end{align}

Differentiating \eqref{eq:app_nll} with respect to $L$ gives
\begin{equation}
\frac{\partial \mathcal{R}}{\partial L} = \frac{1}{1+\exp(-L)} -q_{t,j},
\label{eq:app_nll_first_derivative}
\end{equation}
and
\begin{equation}
\frac{\partial^2 \mathcal{R}}{\partial L^2} = \frac{\exp(-L)}{\left(1+\exp(-L)\right)^2}>0.
\label{eq:app_nll_second_derivative}
\end{equation}
The positive second derivative shows that the expected NLL is strictly convex in $L$. Its unconstrained minimum satisfies
\begin{equation}
\frac{1}{1+\exp(-L^\star)} = q_{t,j},
\end{equation}
which gives
\begin{equation}
L^\star = \log\frac{q_{t,j}}{1-q_{t,j}}.
\label{eq:app_optimal_logit}
\end{equation}
When $|L|\le L_{\max}$ is imposed, the minimizer is the projection of $L^\star$ onto the interval $[-L_{\max},L_{\max}]$. This gives exactly the clipped probabilistic prior in \eqref{eq:prob_prior_optimal}. The KL divergence expression in \eqref{eq:prior_kl_regret} provides the equivalent probability-domain interpretation: the excess expected NLL is nonnegative and vanishes when the assigned probability matches $q_{t,j}$.

\subsection{Variance of the Signed Prior Contribution}

To examine the effect of prior magnitude under prediction uncertainty, define the signed prior contribution as
\begin{equation}
Z_{t,j}(L) = (2B_{t,j}-1)L.
\label{eq:app_signed_prior}
\end{equation}
A positive value indicates that the prior LLR supports the realized bit, whereas a negative value indicates that it supports the opposite bit.
Conditioned on $\mathcal{H}_t$, the mean of $Z_{t,j}(L)$ is
\begin{equation}
\mathbb{E}\left[Z_{t,j}(L)\mid\mathcal{H}_t\right]=(2q_{t,j}-1)L.
\label{eq:app_signed_mean}
\end{equation}
Since $Z_{t,j}^2(L)=L^2$ for either bit value,
\begin{equation}
\mathbb{E}\left[Z_{t,j}^2(L)\mid\mathcal{H}_t\right]=L^2.
\end{equation}
The conditional variance is
\begin{align}
\operatorname{Var}\left(Z_{t,j}(L)\mid\mathcal{H}_t\right)&=
L^2-(2q_{t,j}-1)^2L^2\nonumber\\
&=4q_{t,j}(1-q_{t,j})L^2.
\label{eq:app_signed_variance}
\end{align}
For the hard prior in \eqref{eq:hard_prior}, $|L_{t,j}^{\mathrm{hard}}|=L_{\max}$, and
\begin{equation}
\operatorname{Var}\left(Z_{t,j}(L_{t,j}^{\mathrm{hard}})\mid\mathcal{H}_t\right)
=4q_{t,j}(1-q_{t,j})L_{\max}^2.
\label{eq:app_hard_variance}
\end{equation}
The probabilistic prior in \eqref{eq:prob_prior_optimal} satisfies $|L_{t,j}^{\mathrm{prob}}|\le L_{\max}$. Therefore, its conditional variance satisfies \begin{align}
&\operatorname{Var}\left(Z_{t,j}(L_{t,j}^{\mathrm{prob}})\mid\mathcal{H}_t\right)\nonumber\\
&\qquad = 4q_{t,j}(1-q_{t,j}) \left(L_{t,j}^{\mathrm{prob}}\right)^2\nonumber\\
&\qquad \le 4q_{t,j}(1-q_{t,j})L_{\max}^2.
\label{eq:app_prob_variance}
\end{align}
The variance comparison shows that the probabilistic prior reduces the strength of the signed prior contribution when the prediction is uncertain. Its conditional variance is strictly smaller than that of the hard prior whenever the probabilistic LLR magnitude is below $L_{\max}$.

The variance reduction does not establish that the probabilistic prior is preferable. The relevant condition follows from the expected-NLL optimality derived above. Consider the favorable case in which the hard point prediction selects the more likely bit,
\begin{equation}
\hat B_{t,j} = \mathbbm{1}\{q_{t,j}\ge0.5\}.
\end{equation}
When
\begin{equation}
\left| \log\frac{q_{t,j}}{1-q_{t,j}} \right|<L_{\max},
\label{eq:app_unsaturated_condition}
\end{equation}
the calibrated optimum lies strictly inside the allowed LLR interval. The probabilistic prior uses this interior optimum, whereas the hard prior uses the boundary value $\pm L_{\max}$. The strict convexity of the expected NLL then gives
\begin{equation}
\mathcal{R}
\left(
r(L_{t,j}^{\mathrm{prob}});q_{t,j}
\right)
<
\mathcal{R}
\left(
r(L_{t,j}^{\mathrm{hard}});q_{t,j}
\right).
\label{eq:app_prob_better_condition}
\end{equation}
A hard prediction with the opposite sign produces an even larger mismatch from the calibrated optimum.

In the highly uncertain regime, $q_{t,j}$ approaches $0.5$ and $L_{t,j}^{\mathrm{prob}}$ approaches zero. The probabilistic prior contributes little information when the bit value cannot be predicted reliably, while the hard prior retains magnitude $L_{\max}$.

For highly reliable predictions satisfying
\begin{equation}
\left|
\log\frac{q_{t,j}}{1-q_{t,j}}
\right|
\ge L_{\max},
\end{equation}
the probabilistic prior reaches the clipping bound. If the hard prediction has the same direction, both priors use the same magnitude $L_{\max}$. Strong prior information is preserved when the predicted bit is sufficiently reliable.
The advantage of the probabilistic construction is concentrated in the uncertain regime, where a hard point prior overstates the available prediction confidence. This analysis concerns the quality of the prior information and does not guarantee a lower LDPC decoding error probability for every channel realization.

\bibliographystyle{IEEEtran}
\bibliography{refs}

\end{document}